\documentclass[10pt,letterpaper]{article}
\usepackage[top=0.85in,left=2.75in,footskip=0.75in]{geometry}

\usepackage{amsmath,amssymb}

\usepackage{enumerate}

\usepackage{changepage}

\usepackage{textcomp,marvosym}

\usepackage{cite}

\usepackage{pdfpages}

\usepackage{nameref,hyperref}

\hypersetup{
    colorlinks = true,
    citecolor=blue,
    linkcolor=blue,
    urlcolor=blue,
}

\usepackage[right]{lineno}

\usepackage[nopatch=eqnum]{microtype}
\DisableLigatures[f]{encoding = *, family = * }

\usepackage[table]{xcolor}

\usepackage{array}

\usepackage{booktabs}

\usepackage{commath}

\usepackage{float}

\usepackage{tikz}
\usetikzlibrary{bayesnet}
\usetikzlibrary{arrows}

\usepackage{algorithm}
\usepackage{algpseudocode}
\usepackage{enumitem}

\newcolumntype{+}{!{\vrule width 2pt}}

\newlength\savedwidth

\raggedright
\usepackage[aboveskip=1pt,labelfont=bf,labelsep=period,justification=raggedright,singlelinecheck=off]{caption}

\makeatletter
\renewcommand{\@biblabel}[1]{\quad#1.}
\makeatother

\usepackage{lastpage,fancyhdr,graphicx}
\usepackage{epstopdf}
\fancyheadoffset[L]{2.25in}
\fancyfootoffset[L]{2.25in}
\newcommand{\ddt}[1]{\frac{\dif{#1}}{\dif{t}}}

\newcommand{\given}{\,|\,}

\usepackage{setspace} 
\newcommand{\response}[1]{{#1}}

\usepackage{ifthen}
\newboolean{includeimages}
\setboolean{includeimages}{true}  % set to true for preprints
\begin{document}
\vspace*{0.2in}

\begin{flushleft}
    {\Large
        \textbf\newline{Random time-shift approximation enables hierarchical Bayesian inference of mechanistic within-host viral dynamics models on large datasets} % Please use "sentence case" for title and headings (capitalize only the first word in a title (or heading), the first word in a subtitle (or subheading), and any proper nouns).
    }
    \newline
    \\
    Dylan J. Morris\textsuperscript{1*},
    Lauren Kennedy\textsuperscript{1} and
    Andrew J. Black\textsuperscript{1}
    \\
    \bigskip
    \textbf{1} School of Computer and Mathematical Sciences, University of Adelaide, Adelaide, SA, Australia
    \\
    \bigskip

    % Insert additional author notes using the symbols described below. Insert symbol callouts after author names as necessary.
    % 
    % Remove or comment out the author notes below if they aren't used.
    %
    % Primary Equal Contribution Note
    % \Yinyang These authors contributed equally to this work.

    % Additional Equal Contribution Note
    % Also use this double-dagger symbol for special authorship notes, such as senior authorship.
    % \ddag These authors also contributed equally to this work.

    % \textcurrency Current Address: Dept/Program/Center, Institution Name, City, State, Country % change symbol to "\textcurrency a" if more than one current address note
    % \textcurrency b Insert second current address 
    % \textcurrency c Insert third current address

    % \dag Deceased

    % Group/Consortium Author Note
    % \textpilcrow Membership list can be found in the Acknowledgments section.

    % Use the asterisk to denote corresponding authorship and provide email address in note below.
    * \hyperlink{dylan.morris@adelaide.edu.au}{dylan.morris@adelaide.edu.au}

\end{flushleft}
\section*{Abstract}

Mechanistic mathematical models of within-host viral dynamics are tools for understanding how a virus' biology and its interaction with the immune system shape the infectivity of a host. The biology of the process is encoded by the structure and parameters of the model that can be inferred statistically by fitting to viral load data. 
The main drawback of mechanistic models is that this inference is computationally expensive because the model must be repeatedly solved. This limits the size of the datasets that can be considered or the complexity of the models fitted.
In this paper we develop a much cheaper inference method \response{for this class of models} by implementing a novel approximation of the model dynamics that uses a combination of random and deterministic processes.
This approximation also properly accounts for process noise early in the infection when cell and virion numbers are small, which is important for the viral dynamics but often overlooked. 
Our method runs on a consumer laptop and is fast enough to facilitate a full hierarchical Bayesian treatment of the problem with sharing of information to allow for individual level parameter differences. 
We apply our method to simulated datasets and a reanalysis of COVID-19 monitoring data in an National Basketball Association cohort of 163 individuals.

% Please keep the Author Summary between 150 and 200 words
% Use first person. PLOS ONE authors please skip this step. 
% Author Summary not valid for PLOS ONE submissions.   
\section*{Author summary}
% Please keep the Author Summary between 150 and 200 words. Use first person. 

Understanding how viruses reproduce within an infected host is crucial for predicting disease progression and evaluating treatments. One way to study this is by using mathematical models that describe how viruses reproduce in the body, such as in the respiratory tract. These models help understand key biological processes, but they require significant computing power, making it difficult to analyse data from large groups of individuals.
In this study we developed a new, faster, method for analysing viral load data. Our approach combines randomness, to account for chance events early in infection, with a more predictable model once the infection is well established. This allows us to speed up calculations while capturing key biological processes and maintaining a high degree of accuracy. Our method is efficient enough to run on a laptop, making it possible to study large datasets using a full statistical approach that accounts for individual differences.
We tested our method on 50 simulated datasets and applied it to real-world data from 163 individuals in a COVID-19 monitoring study. Our results show that the model can accurately estimate key parameters and track viral load dynamics over time.

% TODO: turn off line numbers
% \linenumbers
\nolinenumbers

% Use "Eq" instead of "Equation" for equation citations.
\section{Introduction}\label{sec:introduction}

Within-host viral dynamics (WHVD) models describe the progression of viral levels within an individual post-infection \cite{perelsonModellingViralImmune2002,baccamKineticsInfluenzaVirus2006,caniniViralKineticModeling2014}. Depending on their complexity, these models can capture key biological processes such as viral replication, immune response, and clearance, or focus on summarising features like peak timing and growth/decline rates \cite{perelsonModellingViralImmune2002,baccamKineticsInfluenzaVirus2006,caniniViralKineticModeling2014,kisslerViralDynamicsAcute2021}.
In a typical acute infection, the amount of virus in an individual---referred to as the viral load (VL)---grows exponentially before peaking (due to the immune response and/or the carrying capacity of the respiratory tract) and then declining \cite{zitzmannHowRobustAre2024}.
Modelling this progression of virus inside a host can provide insight into important features like when an individual's infectiousness (VL) peaks, differences between variants of a disease, and the vaccination status of individuals \cite{kisslerstephenm.ViralDynamicsSARSCoV22021}, all of which are drivers of between-host (or transmission) dynamics \cite{serenaReviewMultilevelModeling2023,keDailyLongitudinalSampling2022,neantModelingSARSCoV2Viral2021} and are
key in informing health policies \cite{ashcroftTesttraceisolatequarantineTTIQIntervention2022}.

Mathematical models of WHVDs are broadly either phenomenological or mechanistic \cite{zitzmannHowRobustAre2024,challengerModellingUpperRespiratory2022,kisslerViralDynamicsAcute2021}.
Phenomenological models are typically regression models that are able to capture key observed macroscopic (measurable) quantities like growth and decline rates as well as the timing of the peak of infection \cite{kisslerViralDynamicsAcute2021,challengerModellingUpperRespiratory2022}.
Such models can provide a good fit to data and predictions of the viral load trajectory but do not yield any understanding of the underlying biology of the system.
Mechanistic modelling overcomes this by constructing models that explicitly track the populations of cells and virus in a region of the body
\cite{keVivoKineticsSARSCoV22021,zitzmannHowRobustAre2024,ciupeInhostModeling2017,liModellingEffectMUC12021,baiProbabilityMajorInfection2019}. 
Typically such a model is specified in terms of the biological interactions of the lower level entities from which solutions for the macroscopic viral load can be derived or simulated.
The major drawback of such models is that they are much more expensive to fit to data because they do not have simple parametric solutions. One of the main contributions of our work here is to develop a radically faster inference approach, such that fitting complicated mechanistic models is now possible on basic consumer hardware.

Our second main contribution is the ability to fit a fully stochastic model rather than just a deterministic one. To put this in context, mechanistic models are themselves broadly either stochastic or deterministic. Deterministic models are much cheaper to solve, but stochastic models 
capture random variation from when the populations are initially small (i.e. early in the infection) that can still have an impact on the time for the viral load to peak \cite{barbourEscapeBoundaryMarkov2015,morrisComputationRandomTimeshift2024}---ignoring this effect potentially biases inferences. 
The additional complexity of stochastic models has lead---with one exception \cite{germanoHybridFrameworkCompartmental2024}---to most inference approaches ignoring early time noise and only fitting simple deterministic models for the mean populations \cite{challengerModellingUpperRespiratory2022,keDailyLongitudinalSampling2022,zitzmannHowRobustAre2024}. In this work we leverage the large scales over which the VL changes (from $10^0$ to $10^8$ virons per $\mu$l of plasma) to derive a novel approximation for the solutions of a fully stochastic WHVD model. 
This facilitates our first contribution by allowing the derivation of a fast and accurate approximation to the likelihood for VL time-series data. It also properly takes account of the early stochasticity in the process that is required for performing unbiased inference.

Our method is fast enough to fit a full hierarchical Bayesian model to datasets of hundreds of individuals, such as the COVID-19 testing dataset collected from National Basketball Association (NBA) players during the 2020--2021 seasons \cite{kisslerstephenm.ViralDynamicsSARSCoV22021,kisslerViralDynamicsAcute2021}. This dataset is among the most complete and detailed publicly available, featuring longitudinal measurements of individuals' VLs over time using common data formats found in the literature \cite{zitzmannHowRobustAre2024,challengerModellingUpperRespiratory2022,baccamKineticsInfluenzaVirus2006,saenzDynamicsInfluenzaVirus2010,hayEstimatingEpidemiologicDynamics2021}.
In contrast to others \cite{kisslerViralDynamicsAcute2021,kisslerstephenm.ViralDynamicsSARSCoV22021,zitzmannHowRobustAre2024,germanoHybridFrameworkCompartmental2024}, we fit a full stochastic model to the 163 measured individuals and the inference algorithm is run on a consumer laptop in approximately one hour.
The method can produce accurate posterior predictive VL trajectories and parameter estimates of the key within-host parameters (e.g. the within-host basic reproduction number, $R_0$) that are consistent with previous studies of respiratory illnesses \cite{zitzmannHowRobustAre2024,challengerModellingUpperRespiratory2022,baccamKineticsInfluenzaVirus2006,saenzDynamicsInfluenzaVirus2010}.
The hierarchical structure of the model allows us to share information from individuals with sufficient data to those whose data is limited, or noisy, during the growth and/or the decline phases of the infection.
Our method can be easily extended to other datasets and other within-host models (i.e. models that include immune responses \cite{keDailyLongitudinalSampling2022,odakaModelingViralDynamics2021}) and other viral infections and so provides a general tool for fitting to WHVD data.\\

\begin{figure}[H]
    \centering
    \ifthenelse{\boolean{includeimages}}{
    \begin{adjustwidth}{-1.25in}{0in}
     \includegraphics{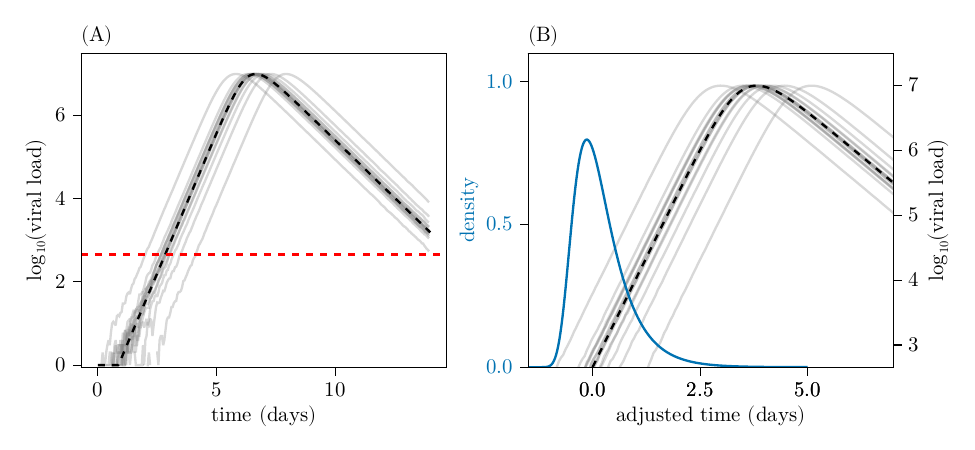}
    \end{adjustwidth}
    }{
        \vspace{0.01mm}
    }
    \caption{
        \textbf{Illustration of the time-shift concept.} (A) Example simulations from the fully stochastic model (grey lines) and the deterministic approximation (black). The detection limit for observations is indicated by the red dashed line. 
        (B) The same simulations as shown in panel (A) but starting the plot from when the simulations cross the detection limit.  
        Translating the deterministic trajectory through time approximates the stochastic solutions on the macroscale. The blue curve shows the probability density of this shift, defined by aligning threshold-crossing times and is equivalent to a shift in the initial conditions of the deterministic model. The models and parameters are described in Sections~\ref{sec:viral_kinetics_model} and \ref{sec:approx_and_macro_dynamics}, with \(\boldsymbol{\theta} = (8, 4, 1.7, 3, 10, 0)\) used in these simulations.
 }\label{fig:time_shift_behaviours}
\end{figure}

The key idea upon which our method builds is that the macroscopic dynamics 
of a fully stochastic WHVD model can be accurately approximated using a deterministic model with a random time-shift applied to the initial conditions \cite{barbourEscapeBoundaryMarkov2015,morrisComputationRandomTimeshift2024}. 
Fig.~\ref{fig:time_shift_behaviours} illustrates this random time-shift idea. In Fig.~\ref{fig:time_shift_behaviours}A, the grey lines show realisations of a stochastic model (defined fully later) and the black line shows the deterministic approximation to the stochastic model. At early times the process is strongly affected by stochasticity due to the small numbers of cells and virions, but this occurs below a detection threshold (indicated by the horizontal red line) where data is not available. 
Once the virion numbers grow large enough, growth becomes exponential, and appears deterministic. The early time stochasticity is reflected on the macroscale (the scale on which measurements can be made, i.e. above the detection threshold) in the time for the virus to peak.
We see that the deterministic model (black line) accurately captures the growth and decline rates as well as the peak height, but not the randomness in the time to peak. But, by time shifting the initial conditions of the deterministic trajectory---keeping the initial numbers fixed---the model can capture the randomness in the time to peak. The blue curve in Fig.~\ref{fig:time_shift_behaviours}B shows the density of this random variable; our previous work details how this can be accurately approximated by a closed form function  that can be calculated for a given model \cite{morrisComputationRandomTimeshift2024}.

Within-host models are an ideal system for an application of this approximation method because the approximation is most accurate where we have data, i.e.~when the populations have grown significantly. During the initial stages of the infection when virus and infected cell numbers are small, the approximation is less accurate, but in this region data is effectively unavailable anyway because there is a detection threshold for measurements (i.e. the data is censored) \cite{zitzmannHowRobustAre2024,challengerModellingUpperRespiratory2022,baccamKineticsInfluenzaVirus2006,saenzDynamicsInfluenzaVirus2010}. The important effect of stochasticity in this early time regime is to shift the time-to-peak, and this is captured accurately by the approximation.

Another way of looking at this approximation is that it allows fast forward simulation of the model without the need for costly stochastic simulation, for example using the Gillespie algorithm \cite{gillespieExactStochasticSimulation1977,gillespieApproximateAcceleratedStochastic2001}. For any given set of parameters, only a \emph{single} solution of the deterministic model and samples from the corresponding time-shift distribution are required.
In many existing studies of this problem the initial viral load is treated as being unobserved and hence a parameter to be inferred \cite{challengerModellingUpperRespiratory2022} or fixed at some value \cite{zitzmannHowRobustAre2024}. 
The time-shift idea facilitates a shift in perspective from inferring the initial variable viral load at a fixed time, to instead inferring a fixed viral load at a random time. This turns out to be a key insight as it allows the derivation of a number of further accurate approximations for the likelihood function that provide large computational speed ups \response{over standard stochastic simulation techniques}, allowing inference on large datasets using a relatively complicated model.

We begin in Section~\ref{sec:methods} by outlining our particular stochastic viral kinetics model and show mathematically how the dynamics can be approximated using the time-shift ideas.
We then provide details of approximations that can be made as a result of how the time-shift is incorporated into the likelihood in order to improve the efficiency of our inference method.
This provides the foundations for a Metropolis-Hastings within Gibbs sampler for sampling from the posterior distribution.
In Section~\ref{sec:results} we apply it to two types of datasets. 
Firstly, multiple synthetic datasets are used to validate the method and quantify its inference capabilities in contexts where the true values of the parameters are known.
We find that our approach is able to disentangle the measurement noise from the process noise and results in parameter estimates that are consistent with the simulated parameters.
Secondly, we fit the model to the NBA COVID-19 dataset \cite{kisslerstephenm.ViralDynamicsSARSCoV22021,kisslerViralDynamicsAcute2021}, where we find that our posterior predictive fits indicate that the method is able to produce a realistic fit to the data and obtain $R_0$ estimates that are consistent with previous studies.
Additionally, we show that the method is able to share information across individuals, to those with few observations and missing data, to estimate their viral dynamics.
All the data and code are available in the following GitHub repository: \href{https://github.com/djmorris7/random_time_shift-within_host_inference}{https://github.com/djmorris7/random\_time\_shift-within\_host\_inference}.

\section{Materials and methods}\label{sec:methods}

\subsection{The viral kinetics model}\label{sec:viral_kinetics_model}

In this work we consider one of the simplest viral kinetics model referred to as the Target-Cell-Limited ($\mathcal{TCL}$) model with an eclipse-phase \cite{baccamKineticsInfluenzaVirus2006,caniniPopulationModelingInfluenza2011,ciupeIdentifiabilityParametersMathematical2022}.
This model captures the basic cell infection cycle and viral production cycles observed during an infection while accounting for an appropriate delay (the eclipse phase) between a cell becoming infected and producing virus.
We choose this simpler model and not something more complex (see \cite{liModellingEffectMUC12021} for an example) as some of the parameters are already unidentifiable from VL data \cite{nguyenAnalysisPracticalIdentifiability2016,ciupeIdentifiabilityParametersMathematical2022} and so adding more complexity without additional data sources only exacerbates this. The method itself can easily be applied to more complex models if required. 
The $\mathcal{TCL}$ model is commonly specified as a set of deterministic ordinary differential equations \cite{baccamKineticsInfluenzaVirus2006,caniniPopulationModelingInfluenza2011,yanExtinctionProbabilityModels2016a} for the total population counts, but here we will instead start with a stochastic version that captures the individual nature of the populations from which we will derive the random time-shift approximation consisting of a deterministic ODE and the time-shift distribution itself.

The $\mathcal{TCL}$ model tracks the numbers of susceptible target cells, $S(t)$, cells in the eclipse phase, $E(t)$, infected cells, $I(t)$, and free virus, $V(t)$, in an effective volume $K$, corresponding to the volume over which the within-host infection process occurs.
The dynamics of these populations are modelled as a continuous-time Markov chain (CTMC) \cite{rossIntroductionProbabilityModels2014,durrettEssentialsStochasticProcesses2016} $\{\boldsymbol{X}(t), t \ge t_0\}$ where $t_0$ is the time at which the infection is initiated, and is unknown, and the state of the system at time $t \ge t_0$ is specified by the vector $\boldsymbol{X}(t) = (S(t), E(t), I(t), V(t))$. Adopting this formulation means that populations are non-negative integers. 
Target cells are uninfected cells that become infected when virus binds to them, a process that follows mass action kinetics \cite{baccamKineticsInfluenzaVirus2006,begonClarificationTransmissionTerms2002}. The instantaneous rate at which target cells become infected is given by $\alpha S(t) V(t) / K$, where $\alpha$ represents the per-virus infection rate of target cells. 
In most formulations of this model 
\cite{baccamKineticsInfluenzaVirus2006,yanExtinctionProbabilityModels2016a,perelsonModellingViralImmune2002,zitzmannHowRobustAre2024}, the parameter $\beta = \alpha / K$ is the parameter of interest. For consistency, we also take $\beta$ as the parameter of interest in our model. 
Following infection, cells enter an eclipse phase, which follows an exponential distribution with mean $1/\sigma$ days, during which infected cells are not yet actively producing virus. Infectious cells are then removed (either by immune clearance or cell death) after an exponentially distributed time with mean $1/\delta$ days. 
While infected, cells produce new virions at rate $\rho$ copies per day, which are cleared at rate $c$ per day, due to immune response or other clearance mechanisms.
The state transitions and rates of the model are summarised in Table~\ref{tbl:tcl_model_rates}.
\begin{table}[!htb]
    \centering
    \begin{tabular}{llc}
        \toprule
        Event & \(\Delta \boldsymbol{X}\) & Rate \\ 
        \midrule
        Infection & \((-1,\ +1,\ 0,\ 0)\) & \(\beta S(t) V(t)\) \\
        Cell becomes infectious & \((0,\ -1,\ +1,\ 0)\) & \(k E(t)\) \\
        Infectious cell recovery & \((0,\ 0,\ -1,\ 0)\) & \(\delta I(t)\) \\
        Production of virus & \((0,\ 0,\ 0,\ +1)\) & \(\rho V(t)\) \\
        Clearance of virus & \((0,\ 0,\ 0,\ -1)\) & \(c V(t)\) \\
        \bottomrule
    \end{tabular}
    \caption{\textbf{Rates for the CTMC viral dynamics model.} Each row gives an event type, the associated change in state \(\boldsymbol{X}(t)\), and the rate at which the event occurs in the \(\mathcal{TCL}\) model.}
    \label{tbl:tcl_model_rates}
\end{table}

There is a choice to be made regarding the initial condition, $\boldsymbol{X}(0)$. The infection is triggered by the entry of virus into the system, so ideally, the initial viral load, $V(0)$, would be set. However, this introduces complexity regarding the amount of virus required to cause an infection \cite{keVivoKineticsSARSCoV22021, smithInfluenzaVirusInfection2018}. 
Smith et al.~\cite{smithInfluenzaVirusInfection2018} show that the dynamics of the model are not significantly altered by taking the initial condition as a single infected cell, as any initial viral particles must initiate an infection by infecting one or more cells (at a rapid rate) before being cleared. 
Throughout this work we assume that $S(0) = 8\times10^7$ based on Zitzmann et al.~\cite{zitzmannHowRobustAre2024}.
Hence, throughout this work, we consider viral infections triggered by a single infected (but not actively infectious) cell, i.e., the initial condition is $\boldsymbol{X}(0) = (S(0) - 1, 1, 0, 0)$. The methods presented here can be extended to handle more complex initial conditions, and extensions of this are discussed in Section~\ref{sec:discussion}.

A pivotal quantity in such models is the basic reproduction number $R_0$, which is defined as the average number of secondary infected cells that are produced by an initially infected cell when the target cell population is fully susceptible \cite{diekmannDefinitionComputationBasic1990}.
The importance of this quantity is that it is a measure of the virus' ability to invade the population of susceptible cells and result in an acute infection \cite{vandendriesscheReproductionNumbersSubthreshold2002}.
If $R_0 \le 1$ then the virus will die out before infecting a large number of cells.
If $R_0 > 1$ then there is a chance that the virus will invade the cell population and result in an acute infection.
The basic reproduction number for this model can be computed through the next generation matrix method \cite{diekmannConstructionNextgenerationMatrices2010} and is given by \cite{zitzmannHowRobustAre2024},
\begin{equation}\label{eq:R0_def}
    R_0 = \frac{\beta S(0) \rho}{\delta c}.
\end{equation}

\subsection{Random time-shift approximation}\label{sec:approx_and_macro_dynamics}

Here, we show that the macroscopic dynamics of the stochastic model described in Section~\ref{sec:viral_kinetics_model} can be approximated by a deterministic ODE model, which corresponds to standard formulations with an added random time shift to the initial conditions. The deterministic approximation follows from a mean-field approximation of the CTMC model in the limit \( K \to \infty \) (i.e., for large volume). 
In the context of respiratory infections, \( K \) corresponds to the total volume over which the within-host infection process occurs and can therefore be considered ``large'' in the sense that the number of interacting particles (e.g., cells and virions) is sufficiently high. Throughout this work, however, the exact value of \( K \) is unimportant, as it can be absorbed into the model parameters.
Taking the large-volume limit yields a system of coupled ODEs for the cell and virus populations (see Section~1 of \nameref{SI:S1_text} for the derivation). This system governs the mean-field dynamics, denoted \(S_d(t), E_d(t), I_d(t), V_d(t)\), where the subscript \(d\) indicates deterministic solutions. To maintain consistency between the stochastic and deterministic formulations, we again make the substitution \(\beta = \alpha / K\), giving
\begin{align}\label{eq:det_model}
    \begin{split}
        \ddt{S_d(t)} & = -\beta S_d(t) V_d(t),        \\
        \ddt{E_d(t)} & = \beta S_d(t) V_d(t) - k E_d(t), \\
        \ddt{I_d(t)} & = k E_d(t) - \delta I_d(t),    \\
        \ddt{V_d(t)} & = \rho I_d(t) - c V_d(t).
    \end{split}
\end{align}
These equations coincide with the ODEs typically used in purely deterministic modelling approaches \cite{zitzmannHowRobustAre2024,caniniViralKineticModeling2014}.

It is difficult to work directly with the transmission parameter \( \beta \) as it is strongly dependent on the experimental setup in the study of interest (e.g., how viral load samples are collected) and negatively correlated with \( \rho \) \cite{zitzmannHowRobustAre2024}. Therefore, we reparameterise the model in terms of \( R_0 \) and back-transform to obtain \( \beta \) when needed. We collect all the parameters in the vector \( \boldsymbol{\theta} = (R_0, k, \delta, \rho, c, t_0) \).

Denote the solution of Eqs.~\eqref{eq:det_model}, $\boldsymbol{X}_d(t) = (S_d(t), E_d(t), I_d(t), V_d(t))$, then this provides an approximation to the mean dynamics of the full stochastic CTMC $\{\boldsymbol{X}(t), t \ge t_0\}$, but does not capture the effect of the noise during the early-time phase when the numbers of infected cells and virions is low (see Fig.~\ref{fig:time_shift_behaviours} and discussion in Section~\ref{sec:introduction}).
The predominant effect of the stochasticity is to shift the time at which the peak VL occurs and subsequently the time at which the infection is cleared \cite{barbourEscapeBoundaryMarkov2015,curran-sebastianCalculationEpidemicFirst2023,morrisComputationRandomTimeshift2024}.
The work of Barbour et al.~\cite{barbourEscapeBoundaryMarkov2015} formalises the connection between the shifted mean field kinetics and samples of the full stochastic process.
They show that $\{\boldsymbol{X}(t), t \ge t_0\}$ is well approximated on the macroscale, i.e.~the growth and decline rates and peak time and height, by another stochastic process $\{\boldsymbol{X}_d(t + \tau), t \ge t_0\}$ where $\tau := \tau(\boldsymbol{\theta}) \in \{-\infty\} \cup \mathbb{R}$ is a scalar random variable referred to as the \textit{random time-shift} \cite{barbourEscapeBoundaryMarkov2015,morrisComputationRandomTimeshift2024}.

The practical importance of this result is that sample paths of $\{\boldsymbol{X}(t), t \ge t_0\}$ resemble those of the deterministic model, $\boldsymbol{X}_d(t)$, except for a translation in time by the time-shift $\tau$. The case $\tau =-\infty$ corresponds to when the number of virions and infected cells drops to zero, i.e.~extinction of the virus within the host. 
Our interest lies in the modelled VL once $V(t)$ has grown to be measurable, i.e.~outside of the early-time small population regime.
Approximating the stochastic process $\{\boldsymbol{X}(t), t \ge t_0\}$ in this way means we can efficiently generate sample paths $V(t)$ by solving for the mean field kinetics and obtaining the distribution of the time-shift.
The process $\{\boldsymbol{X}_d(t + \tau), t \ge t_0\}$ does not capture the noisy early dynamics (i.e. the actual stochastic fluctuations in the populations when these counts are small) but this is of little consequence as small VLs fall below a detection threshold and are therefore not measurable  \cite{zitzmannHowRobustAre2024,kisslerViralDynamicsAcute2021,challengerModellingUpperRespiratory2022}.

The distribution of the random variable $\tau$ is derived by analysing a branching process approximation to the early time dynamics \cite{morrisComputationRandomTimeshift2024}.
Previous work \cite{morrisComputationRandomTimeshift2024} showed that the density function for $\tau$, conditional on non-extinction, can be accurately approximated by a member of a family of functions with the form
\begin{equation}\label{eq:tau_pdf}
    f(\tau \given \boldsymbol{\theta}) = \frac{p\lambda\mu_W^d}{a^d \Gamma\left(\frac{d}{p}\right)} \exp{\left( -\left( \frac{\mu_W e^{\lambda \tau}}{a} \right)^p + \lambda \tau d \right)}
\end{equation}
with $(a, d, p)$ determining the particular member of the family. 
The quantity $\lambda$, representing the early growth rate of the model, and the time-shift parameters $\boldsymbol{\zeta} = (a, p, d)$ are numerically calculated from the branching process approximation to the CTMC (see \cite{morrisComputationRandomTimeshift2024} for concrete examples). 
The blue curve shown in Fig.~\ref{fig:time_shift_behaviours} is this density evaluated for the parameters given in the caption.
Recall that as $\tau$ is a function of $\boldsymbol{\theta}$, the parameters $\boldsymbol{\zeta}$ are also a function of $\boldsymbol{\theta}$ but that this has been suppressed in the notation for clarity.
One final and important point is that the time-shift is independent of the infection time $t_0$.

Computing the parameters of a single time-shift distribution is not expensive, but when incorporated into an Markov chain Monte Carlo (MCMC) routine, where they must be repeatedly evaluated, it can become a bottleneck. 
To overcome this, we train a neural network (NN) to learn 
the function $g : \mathbb{R}^5 \mapsto \mathbb{R}^3$, mapping $\boldsymbol{\theta}=(R_0, k, \delta, \rho, c)$ to $\boldsymbol{\zeta}=(a,b,p)$.
This approach, known as amortised optimisation \cite{amosTutorialAmortizedOptimization2023,marinoGeneralMethodAmortizing2018,gershmanAmortizedInferenceProbabilistic2014}, replaces repeated optimisation with a learned mapping, significantly decreasing the cost of computation during the running of the algorithm, but at the expense of initially training the network \cite{gershmanAmortizedInferenceProbabilistic2014}. Full details of our implementation are provided in Section 6 of \nameref{SI:S1_text}. In testing we found that this modification to our method accelerates the computation of the time-shift parameters by roughly 200 times.

\subsection{Viral load measurement model}\label{sec:measurement_model}

The data we assume available for each individual is a  time-series of the $\log_{10}$ viral load (log-VL) measurements, denoted $\boldsymbol{y} := (y_1, \dots, y_J)$, where $y_j := y_{t_j}$ is the log-VL at time $t_j$.
In practice the $t_j$ could be determined in a variety of ways depending on the data-collection process. 
For a longitudinal study like that run on the NBA cohort \cite{kisslerViralDynamicsAcute2021}, daily observations are taken over a long period of time and only the start point of the time-series is effectively random. 
In a household study, for example, the first time for each individual in a given household is intrinsically linked to both the household being first tested and could come after individuals are infectious, meaning that the first times are random for each index case \cite{marcatoLearningsAustralianFirst2022}. 
Additionally, the timings of the test between individuals are likely to be non-equidistant and also prone to variability as well.
The data is left-censored at a threshold $\eta$ (i.e. $y_j = \eta$ means that $y_j \le \eta$) which describes the limit of detection of the assay used to measure the viral load \cite{kisslerViralDynamicsAcute2021,challengerModellingUpperRespiratory2022,keVivoKineticsSARSCoV22021}.

We assume that the data for a single individual are noisy measurements of some underlying VL trajectory.
Once more we note that the time-shift, $\tau$, is a random variable that is a function of the VL model parameters, $\boldsymbol{\theta}$ and that $V_d(t, \boldsymbol{\theta})$ now has explicit dependence on the VL model parameters $\boldsymbol{\theta}$.
Inclusion of the time-shift means that the modelled trajectory is effectively translated in time and so the modelled trajectory is delayed or advanced depending on the sign of the time-shift.
In the case where there is a slow start to infection (i.e. a negative time-shift) the modelled trajectory is translated to the right and so is effectively delayed.
To handle inconsistencies with the modelled trajectory from applying the time-shift we simply set the VL over this time to the initial condition.
For the case where the time-shift is positive, the opposite is true and the modelled trajectory is effectively translated to the left. 
In this case we can use the modelled trajectory without changes. 
The only other consideration is that the time $t$ we evaluate the modelled trajectory at must be after the infection time, i.e.~$t \ge t_0$.
This can be summarised by defining
\begin{align}\label{eq:det_to_stoch_map}
    z(t, \tau, \boldsymbol{\theta}) = \begin{cases}
        \log_{10}V_d(t + \tau, \boldsymbol{\theta}), \quad &\textrm{ if } t + \tau > t_0 \\ 
        -\infty, \quad &\textrm{ if } t + \tau \le t_0 \textrm{ or } t \le t_0. 
    \end{cases}
\end{align}
\response{Note that predicted viral loads below one particle (i.e. $V(t) < 1$) were mapped to a small pseudocount ($10^{-5}$) before log-transformation to avoid numerical instabilities.}

Let $\boldsymbol{z}(\tau) := (z_1(\tau), \dots, z_J(\tau))$ denote the modelled viral load at the observation times $(t_1, \dots, t_J)$ for a particular time-shift $\tau$, that is $z_j(\tau) := z_{t_{j}}(\tau) := z(t_j, \tau, \boldsymbol{\theta})$.
Throughout this work we will use the convention that normal distributions are parameterised in terms of mean and standard deviation, i.e.~$\mathcal{N}(\mu, \sigma)$, as this parallels the formulation in the code. 
We assume that the observations are noisy measurements of the modelled trajectory, $y_j = z_j(\tau) + \epsilon_j$ and $\epsilon_j \sim \mathcal{N}(0, \kappa)$  for $j = 1, \dots, J$ and some standard deviation $\kappa$, which is assumed to be unknown.

To provide further clarity on the idea of the time-shift and how this connects to the complete data generation process we now illustrate the idea in the context of a forward simulation framework. 
Algorithm~\ref{alg:forward_sim} details the key steps to simulate an approximate stochastic realisation of the WHVD model given model parameters $\boldsymbol{\theta}$. 
This process assumes some discrete times that the stochastic trajectory is sampled at and a particular value of the scale parameter for the observation noise, $\kappa$.

\begin{algorithm}
    \caption{\textbf{Forward simulations using time-shifts}}\label{alg:forward_sim}
    \hspace*{\algorithmicindent} \textbf{Inputs}: 
    \begin{itemize}[noitemsep, topsep=0pt]
        \item The WHVD model parameters $\boldsymbol{\theta}$.
        \item Observation times $\boldsymbol{t} = (t_1, \dots, t_J)$.
        \item Scale parameter on the observation noise, $\kappa$. 
    \end{itemize}
    
    \begin{algorithmic}[1]
        \State Solve Eqs.~\eqref{eq:det_model} to get the deterministic trajectory $V(t)$ through time. 
        \State Compute the time-shift parameters $\boldsymbol{\zeta} = (a, d, p)$ using methods in Morris et. al~\cite{morrisComputationRandomTimeshift2024}.
        \State Sample $\tau \sim f(\tau \given \boldsymbol{\theta})$ (given by Eq.~\eqref{eq:tau_pdf}).
        \State Shift $V(t)$ via Eq.~\eqref{eq:det_to_stoch_map} to get the approximate stochastic trajectory $z(t, \tau, \boldsymbol{\theta})$. 
        \State \textbf{Return } $z(t, \tau, \boldsymbol{\theta})$.
    \end{algorithmic}
\end{algorithm}

\subsection{Mixed effects model}\label{sec:mixed_effects_model}

In this section we provide the general form (deferring details for priors and the calculation of the likelihood to Sections~\ref{sec:likelihood_approximations} and \ref{sec:priors}, respectively) of our mixed effects model. This includes detailing the data and writing down the posterior distribution. 

We assume a simple experimental setup of a sample of $N$ individuals where $\boldsymbol{\theta}_i$ denotes the VL model parameters for individual $i$ and $\boldsymbol{y}_i = (y_{i, 1}, \dots, y_{i, J_i})$ denotes the time-series of their measured VL, which consists of $J_i$ observations. 
Define $\boldsymbol{\Theta} = \{\boldsymbol{\theta}_1, \dots, \boldsymbol{\theta}_N\}$ and $\mathcal{D} = \{\boldsymbol{y}_1, \dots, \boldsymbol{y}_N\}$ as the collections of parameters and data, respectively.
We assume that the subject-specific parameters, \(\boldsymbol{\theta}_i\), are drawn from a population distribution governed by what we refer to as the shared parameters \(\boldsymbol{\phi}\) (the exact relationships are defined in Section~\ref{sec:priors}).
The joint posterior is given by,  
\begin{equation}\label{eq:joint_posterior_simplified}
    f(\boldsymbol{\Theta}, \boldsymbol{\phi}, \kappa \given \mathcal{D}) \propto  \left[ \prod_{i = 1}^{N}   
    f(\boldsymbol{y}_i 
   \given
   \boldsymbol{\theta}_i, \kappa)f(\boldsymbol{\theta}_i \given \boldsymbol{\phi}) \right] f(\boldsymbol{\phi}) f(\kappa).
\end{equation}
The difficult part of evaluating Eq.~\eqref{eq:joint_posterior_simplified} is in the calculation of the marginal likelihood terms, $f(\boldsymbol{y}_i \given \boldsymbol{\theta}_i, \kappa)$, due to the need to marginalise over the distribution of trajectories.

\subsection{Likelihood approximations}\label{sec:likelihood_approximations}

In this section, we outline the construction of the \textit{marginal likelihood} for an individual, breaking it down into its components: the time-shift and the \textit{path likelihood}, the likelihood function given a viral load (VL) sample path. This approach mirrors the general method used in the literature for constructing the likelihood in purely deterministic models \cite{zitzmannHowRobustAre2024,challengerModellingUpperRespiratory2022,liEnhancedViralInfectivity2023}.

The \textit{marginal likelihood} (likelihood for individual $i$'s data, $\boldsymbol{y}_i$, given the parameters, $\boldsymbol{\theta}_i$) is given by $f(\boldsymbol{y}_i \given \boldsymbol{\theta}_i, \kappa)$.
Since the observation process depends on an unobserved random time-shift \(\tau\), we marginalise over \(\tau\), considering only the non-extinction case
\begin{equation}\label{eq:marginal_likelihood_simple}
    f(\boldsymbol{y}_i \given \boldsymbol{\theta}_i, \kappa) = \int f(\boldsymbol{y}_i \given \tau, \boldsymbol{\theta}_i, \kappa) f(\tau \given \boldsymbol{\theta}_i) \dif{\tau}.
\end{equation}
The conditional likelihood given \(\tau\) depends on the system's trajectory \( \boldsymbol{z}_i(\tau) \), which is deterministic for a given \(\tau\) and \(\boldsymbol{\theta}\), so we can rewrite the marginal likelihood as
\begin{equation}\label{eq:marginal_likelihood}
    f(\boldsymbol{y}_i \given \boldsymbol{\theta}_i, \kappa) = 
    \int f(\boldsymbol{y}_i \given \boldsymbol{z}_i(\tau), \boldsymbol{\theta}_i, \kappa) f(\tau \given \boldsymbol{\theta}_i) \dif{\tau}.
\end{equation}

Observations \( y_{i,j} \) correspond to viral load (VL) measurements at times \( t_j \), which are assumed to be independent given the model trajectory. The deterministic trajectory \( z_{i,j}(\tau) := z(t_j, \tau, \boldsymbol{\theta}_i) \) represents the modelled VL at time \( t_j \) for individual \( i \). Given \( z_{i,j}(\tau) \), the conditional likelihood of the observed data \( \boldsymbol{y}_i \) is  
\begin{equation}\label{eq:path_likelihood}
    f(\boldsymbol{y}_i \given \boldsymbol{z}_i(\tau), \boldsymbol{\theta}_i, \kappa) = \prod_{j = 1}^{J_i} f(y_{i,j} \given z_{i, j}(\tau), \boldsymbol{\theta}_i, \kappa),
\end{equation}  
where the terms inside the product are given by 
\begin{align*}
    f(y_{i,j} \given z_{i, j}(\tau), \boldsymbol{\theta}_i, \kappa) = \begin{cases}
        \mathcal{N}_{\textrm{PDF}}(y_{i,j} \given z_{i, j}(\tau), \boldsymbol{\theta}_i, \kappa), \quad \textrm{for } y_{i, j} > \eta, \\ 
        \mathcal{N}_{\textrm{CDF}}(\eta \given z_{i, j}(\tau), \boldsymbol{\theta}_i, \kappa), \quad \textrm{for } y_{i, j} \le \eta
    \end{cases}
\end{align*}
where \( \mathcal{N}_{\textrm{PDF}}(y_{i, j} \given z_{i,j}(\tau), \boldsymbol{\theta}, \kappa) \) and \( \mathcal{N}_{\textrm{CDF}}(\eta \given z_{i,j}(\tau), \boldsymbol{\theta}, \kappa) \) are the probability density function, and cumulative distribution function (CDF), respectively, of a \( \mathcal{N}(z_{i, j}(\tau), \kappa) \) random variable. 
The CDF terms account for observations below the detection threshold \cite{challengerModellingUpperRespiratory2022}. 
To distinguish this from other likelihood terms, we refer to Eq.~\eqref{eq:path_likelihood} as the \textit{path-likelihood} for a single individual.
We note here that if the VL model is replaced with the deterministic model without the time-shift, then this is equivalent to the time-shift density collapsing to a point mass at $\tau = 0$. 
Hence, the likelihood is simply $f(\boldsymbol{y}_i \given \boldsymbol{z}_i(0), \boldsymbol{\theta}_i, \kappa)$ as often seen in VL kinetics modelling \cite{keDailyLongitudinalSampling2022,zitzmannHowRobustAre2024,challengerModellingUpperRespiratory2022,liEnhancedViralInfectivity2023}. 

The marginal likelihood, Eq.~\eqref{eq:marginal_likelihood}, can be evaluated by substituting the path-likelihood, Eq.~\eqref{eq:path_likelihood}, and integrating. However, it involves an integral over a function that does not admit an analytical solution, so numerical methods are required. Numerical integration requires repeated computation of the modelled viral load as we integrate over $\tau$, which appears inside the ODE solution (i.e., $\boldsymbol{z}_i(\tau)$). Evaluating the ODE at each integration step, even with interpolation, is computationally expensive.

To address this, we consider a Laplace approximation \cite[Chapter~27]{mackay2003information} of the interior of the integral of the marginal likelihood. 
The approximation is constructed by noting that for fixed $\boldsymbol{\theta}_i$, $\kappa$ and data $\boldsymbol{y}_i$, the marginal likelihood is simply a function of $\tau$. 
Let 
\begin{align}\label{eq:interior_of_marginal}
    h(\tau) &= f(\boldsymbol{y}_i \given \boldsymbol{z}_i(\tau), \boldsymbol{\theta}_i, \kappa) f(\tau \given \boldsymbol{\theta}_i),  
\end{align}  
and let $g(\tau) = \log{h(\tau)}$. 
Then the Laplace approximation to the marginal likelihood (Eq.~\eqref{eq:marginal_likelihood}) is given by (see Section~4 of \nameref{SI:S1_text} for derivation)
\begin{equation}\label{eq:likelihood_laplace_approx}
    f(\boldsymbol{y}_i \given \boldsymbol{\theta}_i, \kappa) \approx \frac{\sqrt{2 \pi} h(\tau_0)}{\sqrt{-g^{\prime\prime}(\tau_0)}},
\end{equation}
noting that $g(\tau)$ is assumed to be concave (implicit in applying the Laplace approximation), which implies \(g^{\prime\prime}(\tau) < 0\) around \(\tau_0\).
The quantity \(\tau_0\), is the maximiser of Eq.~\eqref{eq:interior_of_marginal} and can be estimated efficiently using optimisation. The denominator, \((-g^{\prime\prime}(\tau_0))^{-1/2}\), is the scale term and has a more complicated form but can also be computed through either numerical differentiation or automatic differentiation of the log of Eq.~\eqref{eq:interior_of_marginal}. In this work, all our code is written in the Julia programming language \cite{bezansonJuliaFreshApproach2017}, and the package ForwardDiff.jl \cite{revelsForwardModeAutomaticDifferentiation2016} enables efficient computation of exact derivatives through forward mode automatic differentiation.
In Section 4 of \nameref{SI:S1_text}, we plot profile likelihoods---i.e., the likelihood with all parameters fixed except for one, which is allowed to vary---for a simulated VL trajectory to show the accuracy of the approximation in Eq.~\eqref{eq:likelihood_laplace_approx} versus the exact calculation in Eq.~\eqref{eq:marginal_likelihood}. The profile likelihoods had a maximum (absolute) error of \(<10^{-7}\) at the tested points, which indicates strong agreement between our Laplace approximation and the exact likelihood.

\subsection{Priors}\label{sec:priors}

In this section, we detail our choices of priors and fixed parameters. We center our priors at values consistent with other studies of COVID-19, and use weakly informative priors with higher variance to account for uncertainty in these estimates. Prior predictive checks confirm that the effective priors on individual parameters remain weakly informative. 
Our primary focus in this work is methodological, demonstrating how to incorporate stochasticity into WHVD models and speed up the inference, rather than conducting a thorough analysis of this data. Moreover, because our hierarchical model is fit to a larger number of individuals than in previous studies, the population-level estimates are better informed by the data and less reliant on the prior. While hierarchical models can be sensitive to prior misspecification, the larger sample size here mitigates this concern, as the data dominate the inference.

Due to identifiability concerns, some VL parameters are fixed at values that are well established in the literature. 
We assume that the eclipse phase of newly infected cells, $k_i$, and the clearance rate of virus, $c_i$, are fixed at $k_i = 4 \textrm{ d}^{-1}$ and $c_i = 10 \textrm{ d}^{-1}$ for $i = 1, \dots, N$, respectively \cite{zitzmannHowRobustAre2024}.
The individual-level parameters, $R_{0, i}$, $\delta_i$, and $\rho_i$, are all assumed to be normally distributed (truncated to be non-negative) with unknown mean and scale parameters. Here we detail the hyper-priors on these parameters.
As outlined in Section~\ref{sec:approx_and_macro_dynamics}, we parameterise the model in terms of the basic reproduction number, $R_0$. 
Table~\ref{tbl:R_0_parameters} gives some estimates of $R_0$ for COVID-19 and the dataset size (number of individuals) from the literature.
None of these studies implemented hierarchical models and instead fit each individual time-series. 
\begin{table}[!ht]
    \centering
    \begin{tabular}{cccl}
        \toprule
        Estimate (95\% CI) & Number of individuals      & Reference & Dataset   \\ \midrule
        8.2 (NA) & 25  & \cite{zitzmannHowRobustAre2024} & NBA \\ 
        13.51 (4.5, 45.12) & 9  & \cite{hernandez-vargasInhostMathematicalModelling2020} & Other \\
        14.2 (NA) & 6  & \cite{germanoHybridFrameworkCompartmental2024} & NBA \\ 
        \bottomrule
    \end{tabular}
    \caption{\textbf{Estimated $R_0$ values from published literature.} Estimates and 95\% CI where available of $R_0$ from published models. The number of individuals in the study are also provided. The dataset column indicates what dataset was of interest for that particular study, with NBA denoting a subset of our dataset and other denoting a different dataset.}
    \label{tbl:R_0_parameters}
\end{table}
To capture the range of values in Table~\ref{tbl:R_0_parameters}, we assume that 
$$
\mu_{R_0} \sim \textrm{Gamma}\left( \frac{10}{3}, 3 \right).
$$
This choice results in a mean of 10 with a 95\% quantile interval of $(2.3, 23.3)$, providing a weakly informative prior centered near the commonly reported estimates.

\response{
To parameterise the priors on the mean infectious period, $\mu_\delta$, and the mean viral replication rate, $\mu_\rho$, we draw on previous studies fitting similar models to COVID-19 data (see \cite{ejimaEstimationIncubationPeriod2021,hernandez-vargasInhostMathematicalModelling2020} and the SI of \cite{zitzmannHowRobustAre2024}). We assume
\[
\mu_\delta \sim \text{Gamma}(26, 0.05),
\]
which corresponds to a mean removal rate of $1.3~\mathrm{d^{-1}}$ and a 95\% interval of $(0.85, 1.85)$, consistent with published estimates.
The parameter $\mu_\rho$ is difficult to assign a prior due to identifiability issues similar to those faced by $\beta$ \cite{zitzmannHowRobustAre2024,germanoHybridFrameworkCompartmental2024}. To address this, we assume
\[
\mu_\rho \sim \text{Gamma}(10, 0.3),
\]
with median $2.9~\mathrm{d^{-1}}$ (roughly based on fits in the SI of \cite{zitzmannHowRobustAre2024}) and a 95\% interval of $(1.44, 5.13)$. This choice reflects prior knowledge while remaining flexible enough for the data to inform the estimate. Note that the main-text model in \cite{zitzmannHowRobustAre2024} includes an immune response, leading to a much larger magnitude ($\sim 100$), so direct numerical comparison is not appropriate.
}

The remaining parameters are assigned weakly informative priors that were selected to guide the fitting process but are not taken from the literature. 
We assign Half-Normal priors on $\sigma_{R_0}$, $\sigma_{\delta}$, and $\sigma_{\pi}$
\begin{align*}
    \sigma_{R_0} \sim \mathcal{N}_{+}(0, 3), \\  
    \sigma_\delta \sim \mathcal{N}_{+}(0, 1), \\
    \sigma_{\rho} \sim \mathcal{N}_{+}(0, 3). 
\end{align*}
The hyper-prior for $\sigma_{R_0}$ and $\sigma_{\rho}$ have a larger standard deviation compared with $\sigma_\delta$ as we found in our testing that a $\mathcal{N}_{+}(0, 1)$ over-constrained the fitted estimates which caused excessive shrinkage of the individual level parameter estimates to the population mean. 
The standard deviation of the measurement noise, $\kappa$, is also assigned a weakly informative prior
\begin{equation*}
    \kappa \sim \mathcal{N}_{+}(0, 1).
\end{equation*}
The priors on the infection times are also chosen to be weakly informative, where we only make the assumption that they are most likely to occur prior to the peak VL measurement,
\begin{equation*}
    t_{0_{i}} \sim \textrm{Gumbel}(-7, 3).
\end{equation*}
This choice results in a median time of infection (pre-peak VL) of $-5.9$ days. 
This prior assumes infection occurs most likely before peak VL but allows for post-peak infection times, which may occur in some cases due to measurement variability or delayed detection. 
Typical scenarios where this prior is useful is when the peak VL is a low measurement or all measurements are tightly clustered in time so it's difficult to distinguish whether the peak is early or late on in the infection.
This distribution is consistent with the incubation periods of COVID-19 (pre-Omicron variant) \cite{sunagawaIsolationMaySelect2023} when assuming that peak infectiousness (i.e. peak VL) corresponds to symptom onset.
For ease of notation we will denote the set of all hyper-parameters as 
$$
\boldsymbol{\phi} = (\mu_{R_0}, \mu_{\delta}, \mu_{\rho}, \sigma_{R_0}, \sigma_{\delta}, \sigma_{\rho}).
$$

\subsection{Inference}\label{sec:implementation_details}

The form of Eq.~\eqref{eq:joint_posterior_simplified} suggests a Gibbs sampler that updates the individual-level parameters ($i = 1, \dots, N$) and the shared parameters $(\boldsymbol{\phi}, \kappa)$ via their conditionals.
However, these conditionals are not available in closed form and cannot be sampled from directly, so we employ Metropolis-Hastings updates within a Gibbs sampling framework \cite[Chapter 10]{robertMonteCarloStatistical2004}.
Full algorithmic details of the sampler, including parallelisation strategies, are provided in Section 5 of \nameref{SI:S1_text}.

All analyses in this paper were implemented in the Julia programming language \cite{bezansonJuliaFreshApproach2017}, which provides seamless integration of model specification, automatic differentiation, and inference within a single environment. For solving ordinary differential equations, we used the high-performance solvers provided by OrdinaryDiffEq.jl \cite{rackauckasDifferentialEquationsJlPerformant2017}. The neural network used to approximate the time-shift distributions was implemented and trained using Flux.jl \cite{Flux.jl-2018}. Automatic differentiation was performed using ForwardDiff.jl \cite{revelsForwardModeAutomaticDifferentiation2016}.

\subsection{Datasets}\label{sec:datasets}

\begin{figure}[ht]
    \centering
    \ifthenelse{\boolean{includeimages}}{
    \begin{adjustwidth}{-1.25in}{0in}
    \includegraphics{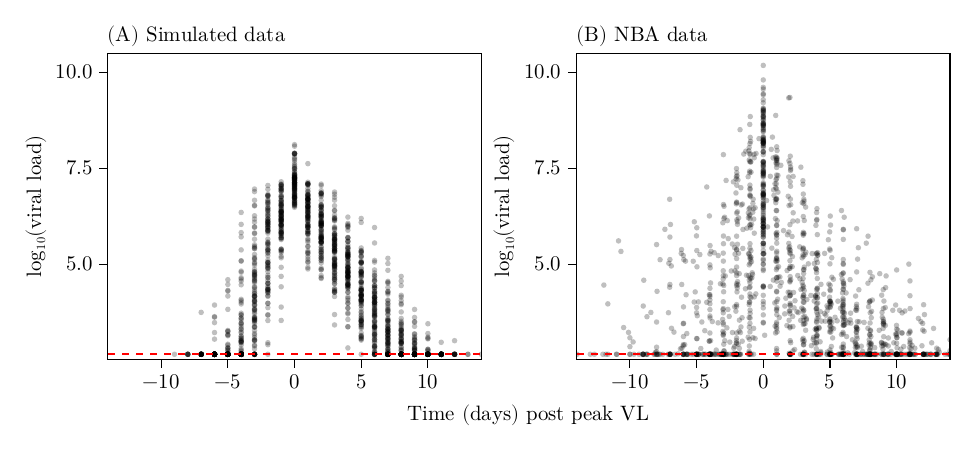}
    \end{adjustwidth}
    }{
        \vspace{0.01mm}
    }
     \vspace{0.01mm}  % invisible but prevents float from collapsing
    \caption{\textbf{Scatter plots of an example simulated dataset and the NBA dataset.} (A) An example synthetic VL dataset generated according to the process in Section~\ref{sec:synth_data}. (B) The NBA VL dataset for the 2020-2021 seasons. In each panel an individual data point corresponds to a single measurement for a single individual. Points are shown as semi-transparent to approximate the ``density" of points, highlighting outliers in panel (B). The red dashed line indicates the limit of detection.}
    \label{fig:data}
\end{figure}

In this analysis, we use a collection of datasets: a published dataset of COVID-19 VL measurements collected during the 2020-2021 NBA season restart, and 200 synthetic datasets for method validation.  
Each synthetic dataset serves as a controlled setting with frequent, low-noise observations and known parameters. The purpose of the simulated datasets is to validate the method can recover parameters in an ideal setting.   
Each dataset consists of \(N = 100\) individual timeseries and are generated using our prior model for the hyperparameters \(\boldsymbol{\phi}\).  
The full data generation process is detailed in Section~\ref{sec:synth_data}, and an example simulated cohort is shown in Fig.~\ref{fig:data}A.

The NBA published COVID-19 monitoring data on players and staff during the 2019-2020 season restart \cite{kisslerViralDynamicsAcute2021,kisslerstephenm.ViralDynamicsSARSCoV22021}.  
Individuals were regularly tested for SARS-CoV-2 from mid-2020 to the season's end, yielding log-transformed viral load measurements relative to peak VL.  
Other studies using mechanistic models \cite{zitzmannHowRobustAre2024,germanoHybridFrameworkCompartmental2024} had to selectively include individuals with frequent pre- and post-peak observations; in our analysis we fit to a much larger set of individuals, many with much poorer data, simultaneously. 
There are 241 individuals in total (68 in \cite{kisslerstephenm.ViralDynamicsSARSCoV22021} and 173 in \cite{kisslerViralDynamicsAcute2021}) but some of these time-series are quite noisy so we performed some pre-processing before fitting. 
Following the approach of Zitzmann et al.~\cite{zitzmannHowRobustAre2024} we excluded individuals with viral loads detected outside a 14-day window around peak VL, as our model does not account for complex behaviours like long-term infections.  

The limit of detection (LOD) in the NBA dataset is 2.658 log copies/mL and frequent testing and individuals experiencing only single infections resulted in many observations (i.e. some individuals having 80\% LOD observations) at this threshold for each individual.
Since it is most likely that individuals are only actually positive for some number of days pre/post the time where they have above LOD observations, we truncated the time series, retaining only the LOD measurements such that above-threshold observations remained between them.  
In some cases, multiple values below the LOD occur before another measurement slightly above the detection limit which complicates fitting.  
To address this, we further truncated time series at three consecutive threshold observations.    
Finally, there were some short time-series with less than 2 observations above the LOD and these were excluded. 
After filtering our dataset contains 163 of the original 241 timeseries. 
Note that we use the same post-processing procedure and LOD for the synthetic dataset.
The NBA data is shown in Fig.~\ref{fig:data}B alongside an example simulated cohort in panel Fig.~\ref{fig:data}A. 
The NBA data (Fig.~\ref{fig:data}B) exhibits visibly higher variance across the entire time horizon compared to the simulated dataset (Fig.~\ref{fig:data}A). This is primarily because the synthetic data were deliberately generated with lower noise to enable more controlled validation of our method.

Our hierarchical model shares information from individuals with high information to improve estimates for those with sparse data, a process known as partial pooling \cite[Chapter~5]{gelmanBayesianDataAnalysis2013}. Previous analyses of this dataset \cite{zitzmannHowRobustAre2024,keVivoKineticsSARSCoV22021,kisslerViralDynamicsAcute2021,kisslerstephenm.ViralDynamicsSARSCoV22021} used deterministic models \cite{zitzmannHowRobustAre2024,keVivoKineticsSARSCoV22021} or piecewise linear functions \cite{keVivoKineticsSARSCoV22021,kisslerViralDynamicsAcute2021,kisslerstephenm.ViralDynamicsSARSCoV22021}, whereas our model explicitly accounts for stochasticity in VL trajectories. Since we include individuals with both high and low information, direct parameter comparisons to the model presented in the 
SI of Zitzmann et al.~\cite{zitzmannHowRobustAre2024} are not possible. However, we can compare our \( R_0 \) estimates to their results and those from other SARS-CoV-2 studies \cite{mccormackModellingViralDynamics2023} as well as other respiratory illnesses \cite{smithInfluenzaVirusInfection2018,mccormackModellingViralDynamics2023}. Additionally, we compare predicted VL trajectories to those of some of the 
SI fits in Zitzmann et al.~\cite{zitzmannHowRobustAre2024}. 
Phenomenological models \cite{challengerModellingUpperRespiratory2022,kisslerViralDynamicsAcute2021} estimate parameters based on piecewise linear models, which are not directly comparable to our mechanistic approach, but our model can estimate the same summary statistics as phenomenological models while also inferring biological quantities like \( R_0 \).  

\subsubsection{Generating synthetic datasets}\label{sec:synth_data}

\response{We generated synthetic viral load (VL) datasets by sampling model parameters from the priors described in Section~\ref{sec:priors}, fixing the shared (hyper-)parameters and drawing individual-level parameters accordingly (e.g. for the basic reproduction numbers, $R_{0,i} \sim \mathcal{N}(\mu_{R_0}, \sigma_{R_0})$ and similarly for $\delta_{i}$ and $\rho_{i}$). The shared parameter means ($\mu_{R_0}$, $\mu_\delta$, $\mu_\rho$) were chosen to align with previous estimates \cite[SI]{zitzmannHowRobustAre2024}, while their corresponding scales ($\sigma_{R_0}$, $\sigma_\delta$, $\sigma_\rho$) were selected heuristically based on comparisons between simulated and NBA datasets. The measurement noise parameter $\kappa$ was set to $0.5$, representing a moderate, ``low-noise" scenario suitable for evaluating model performance under idealised conditions. The full set of hyper-parameter values is given in Table~\ref{tbl:shared_pars_for_synth}.}

\begin{table}[ht]
    \centering
    \begin{tabular}{@{}lc@{}}
        \toprule
        Parameter       & Value   \\ \midrule
        $\mu_{R_0}$     & $8$ \\
        $\sigma_{R_0}$  & $0.5$   \\
        $\mu_\delta$    & $1.3$  \\
        $\sigma_\delta$ & $0.15$   \\
        $\mu_\rho$       & $3$  \\
        $\sigma_\rho$    & $0.25$   \\
        $\kappa$        & $0.5$ \\
        \bottomrule
    \end{tabular}
    \caption{\textbf{The values of the shared parameters used to generate the synthetic datasets.}}
    \label{tbl:shared_pars_for_synth}
\end{table}

\response{For each individual, infection times were drawn from the prior, and corresponding time series were simulated using Algorithm~\ref{alg:forward_sim}. Because variability in the priors can produce unrealistic trajectories (e.g., non-clearing infections or implausible peak VLs), we applied a rejection-sampling procedure to retain only biologically plausible cases. Specifically, we accepted simulations with peak log-VL between 5 and 10 (consistent with the NBA dataset), peak time between 4 and 9 days post-infection, and log-VL below 4.5 at 21 days post-infection to ensure viral clearance. As in the NBA data, a limit of detection (LOD) of 2.658 log copies/mL was imposed and processed as described in Section~\ref{sec:datasets}. The resulting datasets thus represent an idealised setting in which individuals have well-resolved observations across both the growth and decline phases of infection.}

\section{Results}\label{sec:results}

In this section, we apply our method to the datasets described in Section~\ref{sec:datasets}.
We first analyse the synthetic datasets to demonstrate the algorithm’s ability to recover accurate parameter estimates in a setting where the true values are known.
We then apply the method to the NBA dataset to illustrate its utility in a real-world scenario where the true parameter values are unknown.

All simulations and models were implemented in Julia v1.12 \cite{bezansonJuliaFreshApproach2017} and executed on a 2021 MacBook Pro with an M1 chip and 16 GB of RAM. Likelihood computations were parallelised across four high-performance CPU cores, as described in Section 5 of \nameref{SI:S1_text}. Pilot runs were used to tune the proposal distributions for both individual and shared parameters.
Following the pilot runs, we initialised three chains from perturbed versions of the posterior means obtained in the pilot phase.
For each simulated dataset, each chain was run for 100{,}000 iterations, with the first 20{,}000 discarded as burn-in.
\response{For the NBA dataset, each chain was run for 200{,}000 iterations, discarding the first 20{,}000 as burn-in.}
The runs for the NBA dataset (the larger of the two) required approximately 15 minutes per chain, giving a total runtime of about 45 minutes.

Trace plots for all shared parameters and a subset of individual parameters were visually inspected for divergences and showed no evidence of poor mixing across chains in either analysis.
All parameters had $\hat{R} \le 1.01$, indicating good within- and between-chain convergence \cite{vehtariRankNormalizationFoldingLocalization2021}, with minimum effective sample sizes exceeding 500.
The acceptance rates were approximately 0.30 for both the individual (averaged across groups) and shared parameters, within the expected range for well-tuned random walk Metropolis proposals \cite{bedardOptimalAcceptanceRates2008}.

\subsection{Simulation study}\label{sec:simulation_study}

\begin{figure}[!ht]
    \centering
    \ifthenelse{\boolean{includeimages}}{
    \begin{adjustwidth}{-2in}{0in}
        \includegraphics{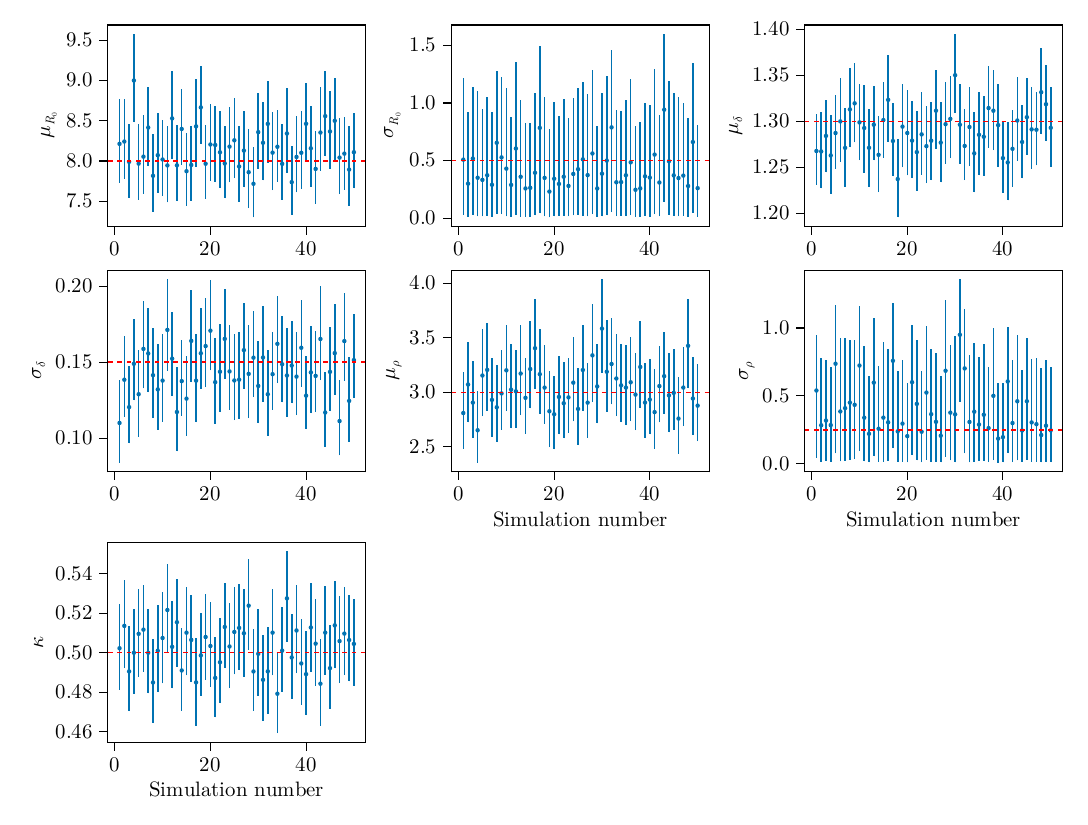}
    \end{adjustwidth}
    }{
        \vspace{0.01mm}
    }
    \caption{\textbf{95\% credible interval plots of the population parameters for 50 of the 200 simulated studies.} Blue lines indicate the 95\% credible interval for a given simulation, with blue points indicating the median. The horizontal red dashed line shows the true value of the parameters used throughout the simulations.}
    \label{fig:sim_coverage_plot}
\end{figure}

\response{Table~\ref{tbl:numerical_coverage_results} reports numerical summaries of the posterior fits (see Section~7 of \nameref{SI:S1_text} for details on each of the scoring rules shown in Table~\ref{tbl:numerical_coverage_results}).
Figure~\ref{fig:sim_coverage_plot} displays the 95\% credible intervals for the population parameters across 50 of the 200 simulated datasets.
In most cases, the true values are contained within the 95\% credible intervals, with the lowest empirical coverage being 0.95.
The credible intervals are narrow relative to the scale of the parameters (see the Interval width column in Table~\ref{tbl:numerical_coverage_results}), indicating that the model recovers the population values with good precision.
Relative bias was computed as the bias divided by the true parameter value, giving a signed percentage difference.
For all parameters, zero lies within the 95\% confidence interval for the relative bias, suggesting that the method is unbiased.
For $\sigma_{R_0}$ and $\sigma_{\rho}$, the average relative biases are $-0.109$ and $0.585$, respectively.
However, Figure~\ref{fig:sim_coverage_plot} shows that the 95\% credible intervals for these variance parameters are much wider (average interval widths of 1.05 and 0.826, respectively), reflecting substantial uncertainty.
This explains why, despite noticeable bias in the posterior means of these two parameters, interval coverage remains adequate.}

\begin{table}[ht]
    \centering
    \begin{tabular}{@{}lccc@{}}
        \toprule
        Parameter       & Coverage & Interval width & Relative bias \\ \midrule
        $\mu_{R_0}$     & $0.95$ & $0.958 \: [0.862, 1.096]$ & $0.016 \: [-0.032, 0.072]$ \\
        $\sigma_{R_0}$  & $1.00$ & $1.05 \: [0.808, 1.233]$ & $-0.109 \: [-0.422, 0.525]$\\
        $\mu_\delta$    & $0.95$ & $0.086 \: [0.078, 0.095]$ & $-0.007 \: [-0.037, 0.025]$ \\
        $\sigma_\delta$ & $0.95$  & $0.056\: [0.05, 0.063]$ & $-0.021 \: [-0.215, 0.146]$ \\
        $\mu_\rho$       & $0.97$ & $0.745\: [0.654, 0.858]$ & $0.020 \: [-0.093, 0.134]$ \\
        $\sigma_\rho$    & $0.98$ & $0.826\: [0.774, 1.443]$ & $0.585 \: [-0.106, 2.067]$ \\
        $\kappa$        & $0.96$ & $0.044\: [0.041, 0.046]$ & $0.006 \: [-0.036, 0.044]$ \\
        \bottomrule
    \end{tabular}
    \caption{
    \textbf{Numerical summary of 200 posterior fits for the shared parameters.}
    Coverage is the proportion of simulations (out of 200) in which the true parameter value lies within the 95\% credible interval. 
    The average interval width is reported across simulations, with a 95\% confidence interval. 
    Relative bias is also reported with its corresponding 95\% confidence interval. 
    \response{See Section~7 of \nameref{SI:S1_text} for further details of the scoring metrics.}
    }
    \label{tbl:numerical_coverage_results}
\end{table}

\begin{figure}[!ht]
    \centering
    \ifthenelse{\boolean{includeimages}}{
    \begin{adjustwidth}{-2.2in}{0in}
        \includegraphics{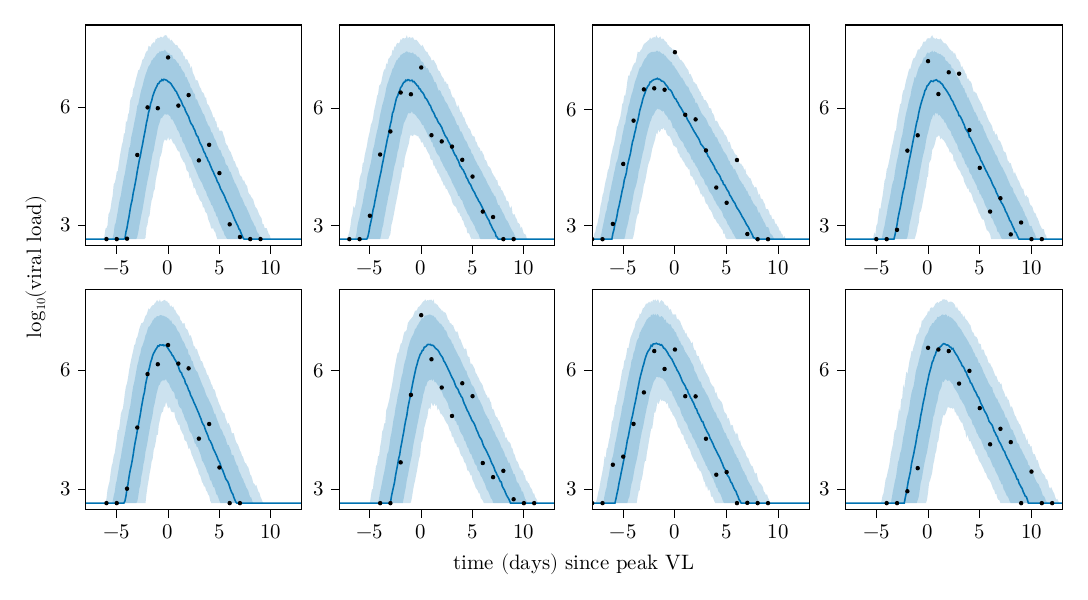}
    \end{adjustwidth}
    }{
        \vspace{0.01mm}
    }
    \caption{\textbf{Posterior predictive fits for a subset of individuals in an example synthetic dataset.} Posterior predictive VL trajectories for a subset of individuals shown in Fig.~\ref{fig:data}. Solid lines represent the median of the posterior predictive fits and the shaded regions represent the 80\% and 95\% prediction intervals. Black dots indicate observed data. }
    \label{fig:trajectories}
\end{figure}

Fig.~\ref{fig:trajectories} shows the posterior predictive fits for some example individuals from the synthetic dataset shown in Fig.~\ref{fig:data}. In all cases, the modelled trajectories track the observed data well, demonstrating that the method can reliably recover the viral dynamics in an ideal scenario. 

\begin{figure}[!ht]
    \centering 
    \ifthenelse{\boolean{includeimages}}{
    \includegraphics{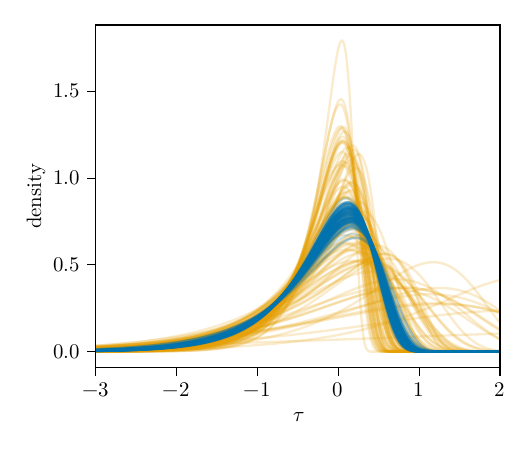}
    }{
        \vspace{0.01mm}
    }
    \caption{\textbf{Time-shift posterior predictive distributions.} Posterior predictive (blue) and prior predictive (orange) time-shifts for a single individual in the synthetic dataset.}
    \label{fig:time_shifts}
\end{figure}

In addition to the trajectories and estimated parameters, by its construction, our model can estimate the posterior predictive time-shift for individuals. Fig.~\ref{fig:time_shifts} compares samples from the the prior (orange) and posterior (blue) time shift distributions for a single individual.
We see that the prior predictive distributions are quite wide and highly varied. Some of the density functions have large amounts of mass concentrated above 1-2 days and have long tails (in both directions). Contrasting this, the posterior predictive time-shift distributions (blue) are more concentrated, with the distributions roughly resembling one another besides the variation arising from the posterior sampled parameters themselves. 

\subsection{NBA study}\label{sec:nba_study}

\begin{figure}[!ht]
    \centering
    \ifthenelse{\boolean{includeimages}}{
    \begin{adjustwidth}{-2in}{0in}
        \includegraphics{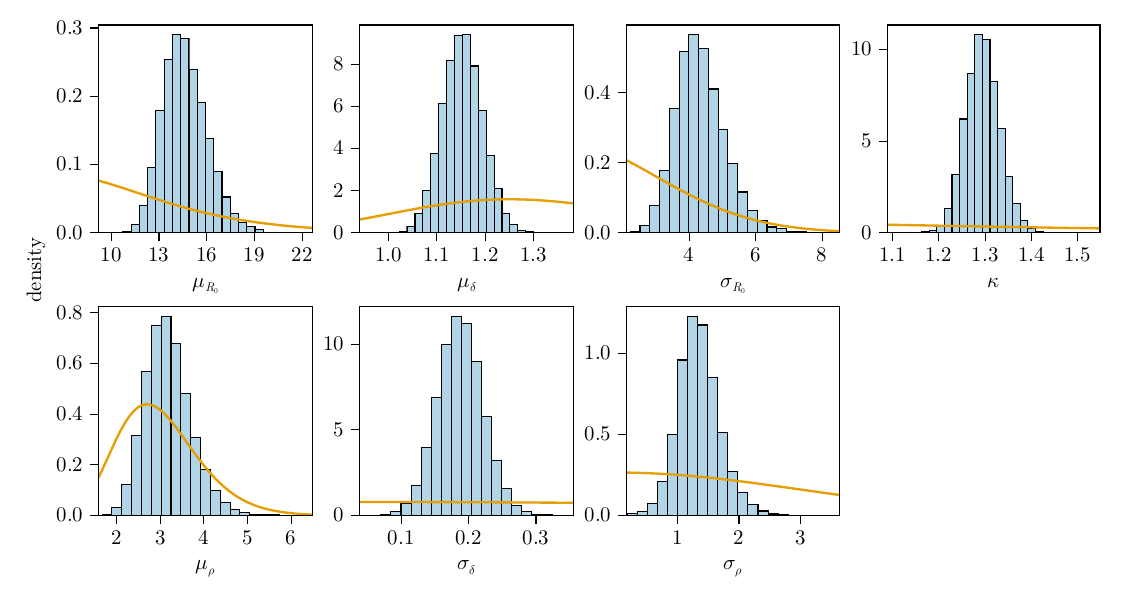}
    \end{adjustwidth}
    }{
        \vspace{0.01mm}
    }
    \caption{\textbf{\response{Histograms} of the marginal posterior distributions for the shared parameters (hyper-parameters and the scale of the measurement noise) for the NBA model fit.} Prior distributions are indicated by orange lines and posteriors are shown as blue histograms.}
    \label{fig:nba_shared_distributions}
\end{figure}

The marginal posteriors for the shared parameters are shown in Fig.~\ref{fig:nba_shared_distributions}. 
We see that there is large movement in the distributions from the prior (orange) to the posteriors (blue) for all parameters.  
The model appears to be able to estimate all parameters including the measurement noise $\kappa$. 
\response{Posterior means and 95\% credible intervals are provided in Table~\ref{tbl:nba_posterior_summary}.}
Our estimates for $\mu_{R_0}$ are consistent with the results in Table~\ref{tbl:R_0_parameters}. 
It should be noted that compared to Zitzmann et al.~\cite[SI]{zitzmannHowRobustAre2024} our population estimate is larger, $14.64$ compared with $8.2$.  
Our $R_0$ estimates are consistent with the value of 14.2 reported in Germano et al.~\cite{germanoHybridFrameworkCompartmental2024} from their stochastic model (which also accounts for immune responses).
Our population estimate for $\mu_{\rho}$ is $3.19$ which is slightly larger than, but still consistent with the estimate of $3.07$ from \cite[SI]{zitzmannHowRobustAre2024}. 
We infer the population mean rate of removal for infectious cells to be $1.15$, slightly lower than $1.3$ in \cite[SI]{zitzmannHowRobustAre2024}. 
There are several possible reasons for the difference in the estimates of $R_0$ (i.e. $\mu_{R_0}$). 
The first being that, as our model is stochastic, larger values of the parameters are required to ensure that the model overcomes extinction than what would be estimated by a deterministic model. 
We are also fitting to a much larger number of individuals (163 vs 25) including those that are unvaccinated and have limited data during the growth and decline phases. 
Additionally, we account for the LOD in our model and make no assumptions as to the time from infection to peak instead allowing our model to infer the infection time.

\begin{table}[!ht]
    \centering
    \begin{tabular}{@{}lc@{}}
        \toprule
        Parameter       & Mean (95\% CrI) \\ \midrule
        $\mu_{R_0}$     & $14.64$ [12.23, 17.60] \\
        $\sigma_{R_0}$  & $4.34$ [3.07, 5.98] \\
        $\mu_{\delta}$  & $1.15$ [1.07, 1.23] \\
        $\sigma_{\delta}$ & $0.19$ [0.12, 0.26] \\
        $\mu_{\rho}$   & $3.19$ [2.28, 4.35] \\
        $\sigma_{\rho}$ & $1.35$ [0.71, 2.10] \\
        $\kappa$        & $1.29$ [1.23, 1.37] \\ 
        \bottomrule
    \end{tabular}
    \caption{
    \textbf{Posterior means and 95\% credible intervals for the shared parameters.}
    }
    \label{tbl:nba_posterior_summary}
\end{table}

\begin{figure}[!ht]
    \centering
    \ifthenelse{\boolean{includeimages}}{
    \begin{adjustwidth}{-2.0in}{0in}
        \includegraphics{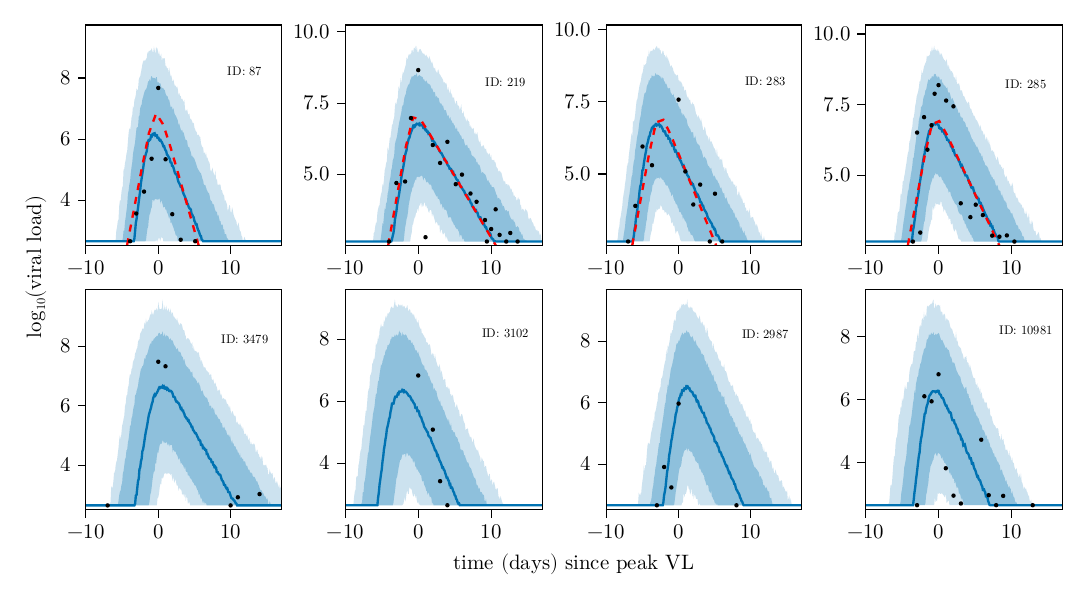}
    \end{adjustwidth}
    }{
        \vspace{0.01mm}
    }
    \caption{\textbf{Posterior predictive fits for a subset of individuals from the NBA dataset.} The first row of plots corresponds to a subset of individuals also presented in the SI of Zitzmann et al.~\cite{zitzmannHowRobustAre2024} for comparison purposes. The second row corresponds to individuals not studied in that work, but that have sparse data that were selected by visual inspection. 
    The observations for each individual are shown in black. 
    The solid lines represent the median of the posterior predictive fits and the shaded regions represent the 80\% and 95\% prediction intervals. The red dashed lines in the first row of plots indicate the deterministic VL trajectories estimated in \cite{zitzmannHowRobustAre2024}. All subplots show the same time period $[-10, 17]$. }
    \label{fig:trajectories_nba_data}
\end{figure}

Fig.~\ref{fig:trajectories_nba_data} shows the posterior predictive VL fits (blue) for a subset of individuals in the NBA dataset. The first row of Fig.~\ref{fig:trajectories_nba_data} corresponds to data for four individuals also analysed in the SI of Zitzmann et al.~\cite{zitzmannHowRobustAre2024} alongside their estimated VL trajectories (red dashed line).
We see strong agreement between our model and the model in the SI of Zitzmann et al.~\cite[SI]{zitzmannHowRobustAre2024} model, with our median predictions closely tracking their estimated trajectories. 
The second row of plots corresponds to individuals (visually) selected for having limited data in either the pre- or post-peak phase. The posterior predictive trajectories track the observed data well.

\section{Discussion}\label{sec:discussion}

We have presented a novel method for fitting mechanistic within-host viral dynamics models that explicitly account for process noise, providing a more realistic representation of viral load (VL) trajectories. Our approach leverages the concept of a random time-shift---first introduced in Barbour et al.~\cite{barbourEscapeBoundaryMarkov2015} and later developed into a practical tool in Morris et al.~\cite{morrisComputationRandomTimeshift2024}---to capture the effects of stochasticity in the WHVDs. Our key contribution in this work is demonstrating how the time-shift distributions can be directly incorporated into the likelihood function, allowing stochasticity to be modelled without the need for computationally expensive stochastic simulations, for example using Gillespie's algorithm \cite{gillespieExactStochasticSimulation1977} or tau-leaping \cite{gillespieApproximateAcceleratedStochastic2001}. Furthermore, this enables efficient inference, making it feasible to fit the model to large datasets using a hierarchical framework. 
\response{Hierarchical approaches have been widely used in phenomenological studies of within-host viral dynamics (WHVDs) \cite{kisslerstephenm.ViralDynamicsSARSCoV22021,challengerModellingUpperRespiratory2022}, but to our knowledge, no stochastic WHVD models have incorporated hierarchical structures.}
Our approach can be viewed as a hybrid method, combining the interpretability and tractability of deterministic models while capturing key elements of stochastic modelling, offering a middle ground between fully stochastic simulation and purely deterministic fitting.

\response{Existing approaches for fitting WHVDs fall into two main categories: simpler phenomenological models, such as piecewise linear functions \cite{kisslerViralDynamicsAcute2021,challengerModellingUpperRespiratory2022,kisslerstephenm.ViralDynamicsSARSCoV22021}, and mechanistic models that assume deterministic dynamics \cite{keVivoKineticsSARSCoV22021,zitzmannHowRobustAre2024,liEnhancedViralInfectivity2023,ciupeIdentifiabilityParametersMathematical2022,ejimaEstimationIncubationPeriod2021,hernandez-vargasInhostMathematicalModelling2020}.
Phenomenological models offer computational advantages, as their likelihoods are easier to evaluate and they integrate well with probabilistic programming frameworks like Stan \cite{stan2024}.
However, they primarily describe viral dynamics in terms of slope parameters (for growth and decline) and a hinge point (for peak VL), limiting their ability to infer biologically meaningful quantities such as the basic reproduction number $R_0$ or production rate $\rho$.
Mechanistic models, by contrast, provide a more realistic representation of infection dynamics but are computationally more demanding, as they require repeatedly solving ODEs or simulating stochastic processes.
Such ODE-based models can be fit using dedicated mixed-effects tools such as Monolix \cite{monolix2024}, which implement efficient frequentist approaches for nonlinear hierarchical models but are less flexible than Bayesian frameworks for incorporating stochastic dynamics or propagating uncertainty in a principled way.
Parameter identifiability also poses a major challenge, particularly when biological parameters influence viral load in overlapping ways (e.g. relationships among $R_0$, the viral production rate, infection timing, and initial conditions) \cite{zitzmannHowRobustAre2024,ciupeIdentifiabilityParametersMathematical2022}.
Although some hybrid stochastic–deterministic models have been developed \cite{rebuliHybridMarkovChain2017,kregerHybridStochasticdeterministicApproach2021,germanoHybridFrameworkCompartmental2024}, they remain computationally intensive and have not been widely applied to large, multi-individual viral kinetics datasets.
Our method is the first to use a stochastic mechanistic model within a hierarchical framework that scales efficiently to datasets with over a hundred individuals.}

We have adopted the simplest viral dynamics model for this work due to parameter identifiability concerns, but the method can easily be applied to other suitable models. These models must satisfy the conditions required for applying the time-shift approximation \cite{morrisComputationRandomTimeshift2024,barbourEscapeBoundaryMarkov2015}, which many biological systems with large populations naturally fulfill \cite{barbourDensityDependentMarkov1980,morrisComputationRandomTimeshift2024,allenPrimerStochasticEpidemic2017}. The main constraint beyond the large population assumption lies in data availability during the pre-exponential growth phase. Systems where early-time dynamics are critical to macroscopic outcomes but where direct observations are limited, such as other within-host models, are particularly well suited for this approach. Examples include the innate response model used in Zitzmann et al.~\cite{zitzmannHowRobustAre2024} and more complex models, such as those examining the role of MUC1 in reducing influenza severity \cite{liModellingEffectMUC12021}.
Conversely, these conditions also define when our method may not be the most appropriate. In cases where high-quality early-growth-phase data is available, exact stochastic methods may be preferable. Epidemic outbreak data, for instance, typically includes detailed observations during the initial growth phase when cases are few and monitoring efforts are robust, making alternative approaches more suitable.

The primary limitation of our method is its complexity compared to simpler approaches in the literature that leverage standard packages like Stan \cite{stan2024}. 
Our framework integrates multiple techniques---neural networks, random time-shift computations, Laplace approximations, and numerical solvers---to fit the model. The neural network plays a critical role in enabling the time-shift approximation by learning a smooth mapping from model parameters to the parameters of the time-shift distribution, effectively circumventing the need for expensive optimisation routines during likelihood evaluation. This learned surrogate is then embedded directly within the likelihood, facilitating fast likelihood computations.
This complexity means the method is not a black box and requires expertise to debug. If performance issues arise, such as MCMC convergence problems, identifying the source of the issue within the pipeline can be challenging.

Increased method complexity often comes with computational challenges, akin to the trade-offs between importance sampling in particle filters and bootstrap filtering methods \cite{blackImportanceSamplingPartially2018,mckinleySimulationbasedBayesianInference2014}. To mitigate these issues, we have structured the method into several modular and easily testable steps. The two main approximations---using a neural network to learn time-shift distribution parameters and employing Laplace approximations---can both be replaced with exact computations (at the cost of efficiency) for validation.
We recommend validating each component separately before applying the full framework. This includes ensuring that the neural network generalizes well beyond its training set, verifying the accuracy of time-shift distributions (e.g., by comparing against simulations from the stochastic model \cite{morrisComputationRandomTimeshift2024}), and using Laplace approximations only when justified (e.g., for normally distributed observation processes). Profile likelihoods can further confirm that the method accurately captures the shape of the exact likelihood, as demonstrated in Section 4 of \nameref{SI:S1_text}.

Another implementation consideration when applying to new models is that the calculation of time-shift distributions are model-specific. Our previous work \cite{morrisComputationRandomTimeshift2024} provides a package for computing the time-shift distributions given the branching process results that can be used for this. Additionally, the neural network must be redesigned and retrained for any new VL kinetics model as the underlying branching process structure is altered. While these challenges are not insurmountable, they are important factors when deciding whether to use this approach.  However, our work here provides a general template for implementation making it substantially easier to extend the method to other models going forward.

Simulation-based validation of viral kinetics models has received relatively little attention in the literature, with most models being fitted directly to data and rarely tested on synthetic datasets. This is likely due to the challenge of generating biologically plausible simulations that capture realistic viral dynamics and experimental designs. In this work, we provide a preliminary validation study using multiple synthetic datasets generated from a single biologically plausible parameter set---drawn from literature estimates and consistent with our real dataset---and show that the true population parameters can be reasonably recovered. Validation and robustness of an inference method has many other dimensions, including model misspecification, varying sample sizes (both within and across individuals), population variability, and measurement error. Exploring inference performance across these factors is an important direction for future work, which would further assess method reliability and contribute to broader validation of viral kinetics models.

A key advantage of our approach is its ability to model heterogeneity in viral trajectories, which is particularly relevant when initial infections involve a small number of cells. 
Beyond within-host modelling, incorporating stochasticity also has implications for between-host transmission studies \cite{wangMultiscaleModelCOVID192022,goyalMultiscaleModellingReveals2022,almoceraMultiscaleModelWithinhost2018,marionModellingUnderstandingPandemics2022}. Small variations in VL trajectories can influence an individual’s infectiousness, affecting transmission dynamics in settings like household studies \cite{marcatoLearningsAustralianFirst2022,blackCharacterisingPandemicSeverity2017,houseInferringRisksCoronavirus2022}. A more realistic representation of within-host variability could lead to improved estimates of key epidemiological parameters, which is an area of ongoing research.

This paper introduces a methodological advancement for working with within-host viral dynamics (WHVDs). Rather than proposing a new model, we focus on developing computational tools that enable efficient Bayesian inference for existing stochastic VL kinetics models. Our interdisciplinary approach---combining applied mathematics, machine learning, and computational statistics---makes it feasible to fit these models to large datasets using biologically motivated, mechanistic, models. By leveraging modern computational techniques, we present a flexible framework that can accommodate the intricacies of viral dynamics and scale efficiently. A key enabler of this work is the use of the Julia programming language \cite{bezansonJuliaFreshApproach2017}, which allows for seamless integration of model specification, automatic differentiation, and inference within a single environment. This capability makes it possible to conduct complex analyses on local hardware, reducing the reliance on dedicated high-performance computing clusters. Overall, the methods introduced here offer a powerful toolset for advancing the study of WHVDs and supporting future research into more realistic and scalable modelling frameworks.

\section*{Supporting information}

% Include only the SI item label in the paragraph heading. Use the \nameref{label} command to cite SI items in the text.
\paragraph*{S1 text.}\label{SI:S1_text}
\textbf{Supplementary text with additional methodological details.} 
This text includes the derivation of the deterministic approximation and the key branching process results underlying the time-shift distributions. 
We provide a full derivation of the Laplace approximation and compare it with the exact calculation using profile likelihoods (\nameref{SI:S1_text}~Fig.~1). 
Further details are given on the model and inference method, including the MCMC routine, as well as the construction of the neural network used for amortised optimisation.

% \section*{Acknowledgments}

\nolinenumbers

%%% TODO: Compile this appropriately as per the journal's requirements before submission.
% \bibliography{My_Library}



\section*{Supplementary material}

# Supporting Information: Random time-shift approximation enables hierarchical Bayesian inference of mechanistic within-host viral dynamics models on large datasets

Dylan J. Morris, Lauren Kennedy, Andrew J. Black

**S1 text: Supporting information**

## 1 Deterministic approximation

Let $\boldsymbol{x}(t) = K^{-1}\boldsymbol{X}(t)$ denote the density process of the CTMC $\{\boldsymbol{X}(t), t \geq t_0\}$. Then reading down column 2 of Table 1 in the main text, and provided $\boldsymbol{a} + \boldsymbol{l} \in \mathcal{S}$ where $\mathcal{S}$ is the state space of the CTMC, $\boldsymbol{a}$ is the current state and $\boldsymbol{l}$ represents the jumps of the process, then the (positive) transition rates can be expressed in terms of the following functions

$$r^{(K)}(\boldsymbol{a}, \boldsymbol{a} + \boldsymbol{l}) = \begin{cases} K\alpha sv, & \text{if } \boldsymbol{\ell} = (-1, 1, 0, 0), \\ Kke, & \text{if } \boldsymbol{\ell} = (0, -1, 1, 0), \\ K\delta i, & \text{if } \boldsymbol{\ell} = (0, 0, -1, 0), \\ K\rho i, & \text{if } \boldsymbol{\ell} = (0, 0, 0, 1), \\ Kcv, & \text{if } \boldsymbol{\ell} = (0, 0, 0, -1). \end{cases} \quad (1)$$

Since the rates can be expressed in this form (referred to as "density dependent"), by Theorem 3.1 of [1], as $K \to \infty$ the process $\boldsymbol{x}(t)$ converges to a deterministic process $\boldsymbol{x}_d(t)$ uniformly in probability over finite time intervals provided $\boldsymbol{x}_d(0) = K^{-1}\boldsymbol{X}(0)$. Letting

$$\boldsymbol{H}(\boldsymbol{x}_d(t)) = \sum_{\boldsymbol{l}} \boldsymbol{l} r(\boldsymbol{a}, \boldsymbol{a} + \boldsymbol{l}),$$

where $r(\boldsymbol{a}, \boldsymbol{a} + \boldsymbol{l}) = K^{-1} r^{(K)}(\boldsymbol{a}, \boldsymbol{a} + \boldsymbol{l})$, then the system of ODEs governing the deterministic approximation is given by

$$\frac{\mathrm{d}\boldsymbol{x}_d(t)}{\mathrm{d}t} = \boldsymbol{H}(\boldsymbol{x}_d(t)). \quad (2)$$

Let $s(t) = S(t)/K$, $e(t) = E(t)/K$, $i(t) = I(t)/K$ and $v(t) = V(t)/K$ denote the densities of the

cell and virus populations of each type, then in the limit as $K \to \infty$ these evolve via a set of coupled ODEs [1] (where the $d$ subscript means these are deterministic)

$$\begin{aligned}
\frac{\mathrm{d}s_d(t)}{\mathrm{d}t} &= -\alpha s_d(t)v(t), \\
\frac{\mathrm{d}e_d(t)}{\mathrm{d}t} &= \alpha s_d(t)v_d(t) - ke_d(t), \\
\frac{\mathrm{d}i_d(t)}{\mathrm{d}t} &= ke_d(t) - \delta i_d(t), \\
\frac{\mathrm{d}v_d(t)}{\mathrm{d}t} &= \rho i_d(t) - cv_d(t).
\end{aligned} \quad (3)$$

We can multiply Eqs. (3) through by the volume $K$ as it is large and fixed to obtain the system of ODEs that govern the approximation for the mean dynamics (Eq. (2) in the main text) [2,3].

## 2 Branching process results

Considering the early stages of infection within a host, the number of target cells, $S$, is approximately constant, i.e. $S \approx S_0$ and hence the rates in Table 1 of the main text are linear. The BPA then tracks the evolution of the state vector $\boldsymbol{X}_b(t) = (E, I, V)$ and the change in event and corresponding rates are given in Table 1.

| $\Delta \boldsymbol{X}$ | Rate |
|---|---|
| $(1,0,0)$ | $\beta S_0 V$ |
| $(-1,1,0)$ | $kE$ |
| $(0,-1,0)$ | $\delta I$ |
| $(0,0,1)$ | $\rho I$ |
| $(0,0,-1)$ | $cV$ |

Table 1: Change in state and rates for the BPA to the $TCL$ model.

To apply the method in [4], we must identify the constant terms from the progeny generating functions which are defined as

$$f_i(\boldsymbol{s}) = \sum_{\boldsymbol{k}} p_i(\boldsymbol{k}) \prod_{j=1}^{m} s_j^{k_j}, \quad \boldsymbol{s} \in [0,1]^m$$

where $p_i(\boldsymbol{k})$ is the probability of an individual of type $i$ having $\boldsymbol{k} = (k_1, \ldots, k_m)$ offspring of each

type [4]. Let $\beta^\star = \beta S_0$, then for the TCL model

$$f_1(\boldsymbol{s}) = \frac{k}{k} s_2 = s_2$$
$$f_2(\boldsymbol{s}) = \frac{\delta + \rho s_2 s_3}{\delta + \rho}$$
$$f_3(\boldsymbol{s}) = \frac{c + \beta^\star s_1 s_3}{c + \beta^\star}.$$

The non-zero constants (see [4] for details) needed for the time-shift computations are

$$\alpha_{12} = k, \quad \beta_{223} = \rho, \quad \beta_{313} = \beta^\star.$$

Note that while this is not used in the current work, the probability of ultimate extinction $q^\star$ is determined by solving for the minimal non-negative solution of the following system [5]

$$f_i(\boldsymbol{q}) = q_i \text{ for } i = 1, 2, 3$$

where the $f_i(\boldsymbol{q})$ are the progeny generating functions. Making the substitution and rearranging the system yields the following system of equations

$$\rho(c + \beta) q_3^2 - (c\delta + \rho\beta) q_3 + c(\delta + \rho) = 0 \tag{4}$$
$$q_1 = \frac{\delta}{\delta + \rho - \rho q_3} \tag{5}$$
$$q_2 = q_1. \tag{6}$$

Since Eq. (4) is quadratic and only in terms of $q_3$, this can be solved analytically and the remaining probabilities determined by substitution.

## 3 Laplace approximation derivation

We evaluated two approaches: (1) constructing a Laplace approximation for the path-likelihood alone and then numerically integrating, or (2) constructing a Laplace approximation for the product of the path-likelihood and the time-shift distribution, and deriving a closed-form expression. The approximation of the path-likelihood alone was initially considered the more appropriate approach, as the path-likelihood is the product of normal PDFs (and few CDF values), and therefore, it is approximately a scaled Normal distribution (note that the scaling means this is a function, not a density). In contrast, the time-shift distribution is skewed (see Fig. 1 in the main text). However, in our testing, we found that these two approaches produced comparable results, with the Laplace approximation of the product of the path-likelihood and time-shift distribution reducing the evaluation time of the likelihood by approximately 30%.

We begin the construction of the Laplace approximation by noting that the data, $\boldsymbol{y}_i$, VL model parameters, $\boldsymbol{\theta}_i$, and $\kappa$ are assumed to be known and fixed. Define the function

$$h(\tau) = \log f(\boldsymbol{y}_i \mid \boldsymbol{z}_i(\tau), \boldsymbol{\theta}_i, \kappa) + \log f(\tau \mid \boldsymbol{\theta}_i).$$

Suppose there is a global maximum

$$\tau_0 = \arg\max_\tau h(\tau),$$

and define

$$H(\tau) = \sqrt{-\frac{1}{h''(\tau)}},$$

noting that $h(\tau)$ is assumed to be concave (implicit in applying the Laplace approximation), which implies $h''(\tau) < 0$ around $\tau_0$, and hence $H(\tau) > 0$.

Taylor expanding $h(\tau)$ to second order, exponentiating, and substituting the result into Eq. (7) in the main text gives the Laplace approximation

$$f(\boldsymbol{y}_i \mid \boldsymbol{\theta}_i, \kappa) \approx f(\boldsymbol{y}_i \mid \boldsymbol{z}_i(\tau_0), \boldsymbol{\theta}_i, \kappa) f(\tau_0 \mid \boldsymbol{\theta}_i) \int \exp\left(-\frac{(\tau - \tau_0)^2}{2H(\tau_0)^2}\right) d\tau.$$

Since the interior of the integrand is a Normal kernel, a closed-form approximation is given by

$$f(\boldsymbol{y}_i \mid \boldsymbol{\theta}_i, \kappa) \approx \sqrt{2\pi} f(\boldsymbol{y}_i \mid \boldsymbol{z}_i(\tau_0), \boldsymbol{\theta}_i, \kappa) f(\tau_0 \mid \boldsymbol{\theta}_i) H(\tau_0). \tag{7}$$

The mean of the Gaussian, $\tau_0$, can be estimated using numerical optimisation. The scale term $H(\tau_0)$ has a more complicated form but can be computed through either numerical differentiation or automatic differentiation. In this work, all our code is written in the Julia programming language [6], and the package ForwardDiff.jl [7] enables efficient computation of exact derivatives through forward mode automatic differentiation for such codes.

We begin the construction of the Laplace approximation by noting that the data, $\boldsymbol{y}_i$, VL model parameters, $\boldsymbol{\theta}_i$, and $\kappa$ are assumed to be known and fixed. Define the functions

$$g(\tau) = f(\boldsymbol{y}_i \mid \boldsymbol{z}_i(\tau), \boldsymbol{\theta}_i, \kappa) f(\tau \mid \boldsymbol{\theta}_i), \quad h(\tau) = \log g(\tau).$$

Suppose there is a global maximum

$$\tau_0 = \arg\max_\tau g(\tau) = \arg\max_\tau h(\tau),$$

and define

$$H(\tau) = \sqrt{-\frac{1}{h''(\tau)}},$$

4

noting that the function is assumed to be concave (implicit in applying the Laplace approximation), which implies $h''(\tau) < 0$ around $\tau_0$, and hence $H(\tau) > 0$. Taylor expanding $h(\tau)$ about $\tau_0$ and noting that $h'(\tau_0) = 0$ (since $\tau_0$ is the maximum), we get

$$h(\tau) \approx h(\tau_0) + \frac{1}{2}h''(\tau_0)(\tau - \tau_0)^2 = h(\tau_0) - \frac{1}{2H(\tau_0)^2}(\tau - \tau_0)^2.$$

Then, we have

$$g(\tau) = \exp\left(h(\tau)\right) \approx g(\tau_0) \exp\left(-\frac{(\tau - \tau_0)^2}{2H(\tau_0)^2}\right).$$

Since $H(\tau_0) > 0$, this is the kernel of a $\mathcal{N}(\tau_0, H(\tau_0))$ random variable scaled by $g(\tau_0) = f(\boldsymbol{y}_i \mid \boldsymbol{z}_i(\tau_0), \boldsymbol{\theta}_i, \kappa)$. This equation is what is substituted into the marginal likelihood and yields Eq. (10) in the main text.

# 4 Comparison of our Laplace approximation versus the exact calculation

Figure 1 shows the profile likelihoods for a given parameter set (for a random individual). These fix all but the parameter of interest and allow that to vary over some interval. The orange dots show our approximation and we can see strong agreement between the approximation and the "exact" likelihood. The maximum (absolute) error computed over the points shown in the plot is $< 1 \times 10^{-7}$.

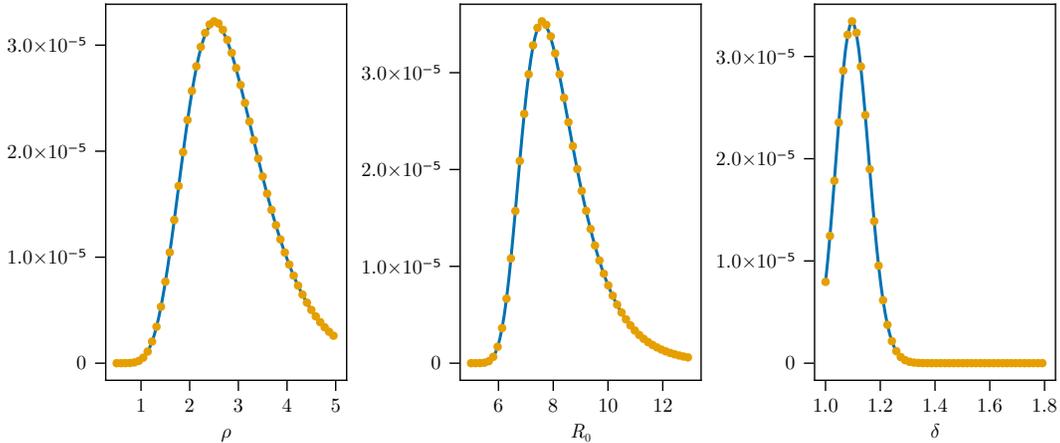

Figure 1: **Profile likelihoods for the exact method and the Laplace approximation.** Profile likelihoods for the two methods: exact given by the solid blue line, Laplace approximation in orange points.

# 5 Model

Relationships between the components in the model are depicted in the plate diagram shown in Fig. 2 where the direction of the arrows dictates dependencies between parameters, hyperparameters and data.

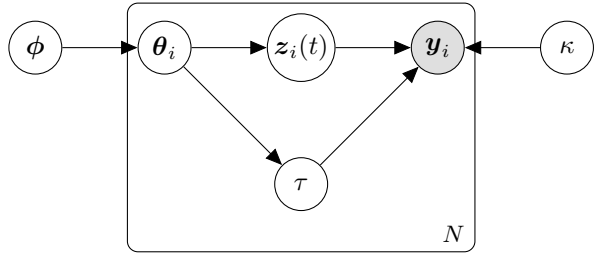

Figure 2: **Plate diagram of the relationship between components in the model.** The plate indicates individual level variables that are indexed by the number of individuals, $N$. The parameters outside the plate, $\phi$ and $\kappa$, are shared across all individuals.

The joint posterior distribution of the model parameters and the shared parameters is given by

$$f(\boldsymbol{\Theta}, \boldsymbol{\phi}, \kappa \,|\, \mathcal{D}) \propto f(\mathcal{D}\,|\,\boldsymbol{\Theta},\kappa)f(\boldsymbol{\Theta}|\boldsymbol{\phi})f(\boldsymbol{\phi})f(\kappa), \tag{8}$$

where $f(\mathcal{D}\,|\,\boldsymbol{\Theta},\kappa)$ is the joint parameter-likelihood, and $f(\boldsymbol{\Theta}|\boldsymbol{\phi})$, $f(\boldsymbol{\phi})$, and $f(\kappa)$ are prior distributions that are detailed in Section 2.6. Since the individuals are conditionally independent given their unique parameters, $\boldsymbol{\theta}_i$, we can factorise the joint-likelihood as

$$f(\mathcal{D}\,|\,\boldsymbol{\Theta},\kappa) = \prod_{i=1}^{N} f(\boldsymbol{y}_i\,|\,\boldsymbol{\theta}_i,\kappa), \tag{9}$$

where $f(\boldsymbol{y}_i\,|\,\boldsymbol{\theta}_i,\kappa)$ is the *marginal likelihood* function for individual $i$, as it is obtained by marginalising over the random time-shifts (see Section 2.5 for details).

## 5.1 Inference

The form of the joint posterior (Eq. (5) in the main text) suggests a Gibbs sampler that draws from the conditional distributions for each of the individuals ($i = 1, \ldots, N$) and the shared parameters ($\boldsymbol{\phi}, \kappa$). However, sampling directly from the conditionals is not possible as the distributions cannot be normalised—or recognised as standard distributions—so we use a Metropolis-Hastings step within the Gibbs sampler. The conditional distributions for the individual parameters, noting that $\boldsymbol{\theta}_i$ are independent of $\boldsymbol{\Theta}_{-i} = \{\boldsymbol{\theta}_1, \ldots, \boldsymbol{\theta}_{i-1}, \boldsymbol{\theta}_{i+1}, \ldots, \boldsymbol{\theta}_N\}$ given $\boldsymbol{\phi}$, are given by

$$f(\boldsymbol{\theta}_i \,|\, \boldsymbol{\phi}, \kappa, \mathcal{D}) \propto f(\boldsymbol{y}_i\,|\,\boldsymbol{\theta}_i,\kappa)f(\boldsymbol{\theta}_i\,|\,\boldsymbol{\phi}).$$

6

The conditional distribution for $\kappa$ and $\phi$ is

$$f(\phi, \kappa \mid \Theta, \mathcal{D}) \propto f(\kappa) f(\phi) \prod_{i=1}^{N} f(\mathbf{y}_i \mid \boldsymbol{\theta}_i, \kappa) f(\boldsymbol{\theta}_i \mid \phi). \tag{10}$$

Let the current draw be $\{\Theta^{(k)}, \phi^{(k)}, \kappa^{(k)}\}$ and we use the notation $\Theta_{a:b} = \{\boldsymbol{\theta}_a, \ldots, \boldsymbol{\theta}_b\}$ for integers $a$ and $b$ ($b > a$). Define the following collections

$$\psi_i^{(k+1)} = \left\{ \Theta_{1:i}^{(k+1)}, \Theta_{i+1:N}^{(k)} \right\}, \quad i = 1, \ldots, N-1,$$
$$\psi_N^{(k+1)} = \left\{ \Theta_{1:N}^{(k+1)} \right\}.$$

We choose a relatively simple multivariate Gaussian proposal distributions centered on the current values of the parameters being drawn, with $N + 1$ separate covariance matrices per step of the method, i.e.

$$\boldsymbol{\theta}_i' \mid \boldsymbol{\theta}_i^{(k)} \sim \mathcal{N}(\boldsymbol{\theta}_i^{(k)}, \boldsymbol{\Sigma}_i), \quad i = 1, \ldots, N,$$
$$(\phi', \kappa') \mid (\phi^{(k)}, \kappa^{(k)}) \sim \mathcal{N}((\phi^{(k)}, \kappa^{(k)}), \boldsymbol{\Sigma}_{N+1}),$$

Blocking the proposals in this way enables the subject specific proposals as well as the proposal for the shared parameters, $\kappa$ and $\phi$, to be tuned separately. The covariance matrices are tuned by performing a sufficiently large pilot run. Denote the PDF of the proposal distributions by $Q(\cdot \mid \cdot)$, where the arguments specify the particular proposal distribution being used. The acceptance probabilities for individual parameters ($i = 1, \ldots, N$) are given by

$$\alpha_{1,i} = \min \left\{ \frac{f(\boldsymbol{\theta}_i' \mid \psi_{i-1}^{(k+1)}, \phi^{(k)}, \kappa^{(k)}, \mathcal{D})}{f(\boldsymbol{\theta}_i^{(k)} \mid \psi_{i-1}^{(k+1)}, \phi^{(k)}, \kappa^{(k)}, \mathcal{D})} \times \frac{Q(\boldsymbol{\theta}_i^{(k)} \mid \boldsymbol{\theta}_i')}{Q(\boldsymbol{\theta}_i' \mid \boldsymbol{\theta}_i^{(k)})}, 1 \right\}, \tag{11}$$

Similarly, the acceptance probability for the shared parameters is

$$\alpha_2 = \min \left\{ \frac{f(\phi', \kappa' \mid \psi_N^{(k+1)}, \mathcal{D})}{f(\phi^{(k)}, \kappa^{(k)} \mid \psi_N^{(k+1)}, \mathcal{D})} \times \frac{Q(\phi^{(k)}, \kappa^{(k)} \mid \phi', \kappa')}{Q(\phi', \kappa' \mid \phi^{(k)}, \kappa^{(k)})}, 1 \right\}. \tag{12}$$

Note that the proposal distributions are symmetric and so the ratio of the proposal distributions in Eq. (11) and Eq. (12) are equal to 1. Algorithm 1 outlines a single step of the Metropolis-within-Gibbs sampler.

### 5.1.1 Parallelisation of the inference method

The structure of the inference algorithm and the form of the conditional distributions allow for straightforward parallelisation. The for loop (lines 1-4 of Algorithm 1) samples the VL model

---

**Algorithm 1** Single step of the Metropolis-within-Gibbs algorithm

**Inputs**:
- Current draw $(\boldsymbol{\Theta}^{(k)}, \boldsymbol{\phi}^{(k)}, \kappa^{(k)})$;
- Conditional distributions; $f(\boldsymbol{\theta}_i \,|\, \boldsymbol{\phi}, \kappa, \mathcal{D})$, $f(\boldsymbol{\phi}, \kappa \,|\, \boldsymbol{\Theta}, \mathcal{D})$;
- Transition kernels; $Q(\boldsymbol{\theta}'_i \,|\, \boldsymbol{\theta}_i)$ for $i = 1, \ldots, N$, $Q(\boldsymbol{\phi}', \kappa' \,|\, \boldsymbol{\phi}, \kappa)$

1: **for** $i = 1, \ldots, N$ **do**
2:     Propose individual parameters $\boldsymbol{\theta}'_i \sim Q(\boldsymbol{\theta}'_i \,|\, \boldsymbol{\theta}_i^{(k)})$.
3:     Compute $\alpha_{1,i}$ (Eq. (11)) and accept proposal with probability $\alpha_{1,i}$.
4: **end for**
5: Propose shared parameters $\boldsymbol{\phi}', \kappa' \sim Q(\boldsymbol{\phi}', \kappa' \,|\, \boldsymbol{\phi}^{(k)}, \kappa^{(k)})$.
6: Compute $\alpha_2$ (Eq. (12)) and accept proposal with probability $\alpha_2$.

---

parameters $\boldsymbol{\theta}_i$ independently of the other individuals given the shared parameters. This means that the sampling of the individual parameters can be parallelised across the individuals. Additionally, parallelisation can also be used when computing the acceptance probability in line 6 as the likelihood contributions from each individual in Eq. (10) (the $\mathcal{L}_i(\boldsymbol{\theta}_i, \kappa)$ terms inside the product) can be computed in parallel conditional on a draw of the hyper-parameters $\boldsymbol{\phi}$.

### 5.1.2 NCP

We use a non-centred parameterisation (NCP) for the individual-level parameters to enhance posterior sampling efficiency [8]. Specifically, the individual-specific parameters $R_{0_i}, \delta_i$, and $\rho_i$ are modelled as

$$R_{0_i} = \mu_{R_0} + \sigma_{R_0} z_{R_{0_i}}, \quad \delta_i = \mu_\delta + \sigma_\delta z_{\delta_i}, \quad \rho_i = \mu_\rho + \sigma_\rho z_{\rho_i}$$

where $z_{R_{0_i}}, z_{\delta_i}, z_{\rho_i} \sim \mathcal{N}(0, 1)$. This parameterisation facilitated more efficient posterior sampling.

## 6 Neural network architecture and details

One particular challenge with using the time-shift approach in a statistical inference framework is that computing the parameters of the distribution is itself a computationally expensive task. Each call of the likelihood requires optimisation to be carried out to determine the parameters of the time-shift distribution [4]. To overcome this performance bottleneck, we utilise a neural network (NN) to learn the parameters of the time-shift distribution, $(a, b, p)$, given VL model parameters, $\boldsymbol{\theta} = (R_0, k, \delta, \rho, c)$. Note the removal of $t_0$ from the VL model parameters as the time-shift distribution does not depend on the infection time. This process is referred to as amortised optimisation [9–11] and is a common technique in machine learning to improve the efficiency of computationally complex algorithms such as Markov chain Monte Carlo [11].

Let $\boldsymbol{\zeta}_i = (a_i, b_i, p_i)$ denote the parameters of the time-shift distribution for some model parameters

$\boldsymbol{\theta}_i$. We generate our dataset by sampling VL model parameters from a hyper-cube that encompasses our model priors. Our previous work [4] is used to determine the time-shift distribution parameters to give the paired data, $\{(\boldsymbol{\theta}_i, \boldsymbol{\zeta}_i)\}_{i=1}^n$. The time-shift distributions are checked for validity purposes since sampling on the hyper-cube does not necessarily produce feasible parameter combinations. The goal here is to learn a function $g : \mathbb{R}^5 \mapsto \mathbb{R}^3$ given by $g(\boldsymbol{\theta}_i) = \boldsymbol{\zeta}_i$ to a high degree of accuracy. The learned function will approximate the mapping from the VL model parameters to the time-shift parameters without the need to perform complex numerical routines. Replacing the numerical routine with this learned function makes the runtime to find the parameters approximately $200\times$ faster.

When generating the training data, we sample parameters on a hyper-cube capturing the range of the independent priors used in the inference. In other words, the we sample the parameters as follows

$$R_0 \sim U(1, 35)$$
$$\delta \sim U(0.01, 10)$$
$$\rho \sim U(0.01, 10).$$

Instead of naively sampling parameters, we check that the growth rate, $\lambda > 0$, and that 95% of the density for $\tau$ is between $[-7, 7]$. The first condition ensures that there is a non-zero probability of the infection going non-extinct before reaching the exponential growth phase. The second condition ensures that long time-shifts, typically about a half of the total viral infection time-span (in our case), are considered un-biological and that time-shifts lying outside range likely suggests an incorrect model specification or poor prior choices. One key detail from an implementation perspective is that we also standardise the inputs (model parameters) and the outputs (time-shift parameters) using standard scaling (i.e. subtract the sample mean for each parameter and divide by the sample variance). In the context of training NN's, this standardisation has the effect of drastically improving the performance and convergence behaviour of optimisers like Adam. We utilise a simple NN with the following architecture:

1. Input layer: 3 neurons (the free parameters of the VL model) with ReLU activation

2. Hidden layer: 64 neurons with ReLU activation

3. Output layer: 3 neurons (the parameters of the time-shift distribution) with soft-plus activation

The input and hidden layer both use the rectified linear unit (ReLU) activation functions. The first of which acts to standardise the input data and then in the output layer we use the soft-plus activation function to ensure non-negativity in the GG parameters. We consider mean-squared-error (MSE) as the loss function to be optimised as this performed well in testing and all the output

parameters for the surrogate distribution are on the same scale. Let

$$\mathcal{A} = \left\{(\boldsymbol{\theta}_i, \boldsymbol{\zeta}_i)\right\}_{i=1}^{n_{\text{train}}} \quad \text{and} \quad \mathcal{B} = \left\{(\boldsymbol{\theta}_i, \boldsymbol{\zeta}_i)\right\}_{i=1}^{n_{\text{test}}}$$

denote the training and validation sets respectively. Let $\mathbb{M}_\Psi(\boldsymbol{\theta})$ denote the current NN model—where we have collected the NN model parameters into $\Psi$—evaluated at mechanistic model parameters $\boldsymbol{\theta}$ which returns the surrogate distribution parameters. Define $\mathbb{M}(\boldsymbol{\theta})_j$ as the $j$th element of this (i.e. the estimated surrogate distribution parameter using the NN), then the loss is given by

$$L(\Psi) = \sum_{i=1}^{n_{\text{train}}} \sum_{j=1}^{3} \left(\mathbb{M}_\Psi(\boldsymbol{\theta_i})_j - \zeta_j\right)^2. \tag{13}$$

We use the ADAM optimiser with defaults as provided by the Julia package Flux.jl. Additionally to improve the generalisability of the model, we perform early stopping. Early stopping uses the validation set and we stop training the model if there is no improvement in the loss (calculated in the same way as the loss in Eq. (13)) on the validation set after 30 epochs. The model is completely implemented and trained in Julia on a CPU (to simplify dissemination of codes) using the Flux.jl package.



\end{document}